\documentstyle[aps,prd,amssymb,12pt,preprint,tighten,epsf]{revtex}

\newcommand{\be}{\begin{equation}}
\newcommand{\ee}{\end{equation}}
\draft
\begin{document}
\title{RELATIVISTIC GRAVITATIONAL COLLAPSE IN NON--COMOVING COORDINATES:
THE POST-QUASISTATIC APPROXIMATION}
\author{L. Herrera$^1$\thanks{Postal
address: Apartado 80793, Caracas 1080A, Venezuela.} \thanks{e-mail:
laherrera@telcel.net.ve}, W.Barreto$^2$\thanks{email:wbarreto@sucre.udo.edu.ve}
, A. Di Prisco$^1$,
N. O. Santos$^{3,4,5}$\thanks{e-mail: nos@cbpf.br; santos@ccr.jussieu.fr}
 \\
{\small $^1$Escuela de F\'{\i}sica, Facultad de Ciencias,}\\
{\small Universidad Central de Venezuela, Caracas, Venezuela.} \\
{\small $^2$ Grupo de Relatividad y Astrof\'\i sica Te\'orica,}\\
{\small Departamento de F\'\i sica, Escuela de Ciencias,}\\
{\small N\'ucleo de Sucre, Universidad de Oriente, Cuman\'a, Venezuela.} \\
{\small $^3$LRM--CNRS/UMR 8540, Universit\'e Pierre et Marie Curie, ERGA,}\\
{\small Bo\^\i te 142, 4 place Jussieu, 75005 Paris Cedex 05, France.}\\
{\small $^4$Laborat\'orio Nacional de Computa\c{c}\~ao Cient\'{\i}fica,}\\
{\small 25651-070 Petr\'opolis--RJ, Brazil.}\\
{\small $^5$ Centro Brasileiro de Pesquisas F\'{\i}sicas, 22290-180
Rio de Janeiro-RJ, Brazil.}\\
}
\maketitle
\begin{abstract}
A general, iterative, method for the description of evolving
self-gravitating relativistic spheres is presented. Modeling is achieved
by the introduction of an ansatz, whose rationale becomes intelligible and
finds full justification within the context of a suitable definition of the
post--quasistatic
approximation. As examples of application of the method we discuss three
models, in the adiabatic case.
\end{abstract}
\pagebreak
\section{Introduction}
The problem of general relativistic gravitational collapse has attracted
the attention of researchers since  the seminal paper by Oppenheimer
and Snyder \cite{Opp}. The motivation for such interest is easily
understood: the gravitational collapse of massive stars represents one of
the few observable phenomena, where general relativity is expected to
play a relevant role. Ever since that work, much was written by
researchers trying to provide models of evolving gravitating spheres.
However this endeavour proved to be difficult and uncertain. Different
kinds of advantages and obstacles appear, depending on the approach adopted
for the modeling.

Thus, numerical methods (see \cite{Lehner} and references therein)
are enabling researchers to investigate systems which are extremely
difficult to handle analytically. In the case of General Relativity,
numerical models have proved valuable for investigations of strong
field scenarios and have been crucial to reveal unexpected phenomena
\cite{choptuik}. Even specific difficulties associated with numerical
solutions of partial differential equations in presence of shocks are
being overpassed
 \cite{font}. By these days what seems to be the main limitation for
 numerical relativity is the computational demands for 3D evolution,
 prohibitive in some cases \cite{winicour}. Nevertheless, purely
numerical solutions usually hinder to catch general, qualitative, aspects
of the process.

On the other hand, analytical solutions although more suitable for a
general discussion (see \cite{Bonnor} and references therein), are
found, either for too simplistic equations of state and/or under additional
heuristic assumptions whose justification is usually uncertain.

Therefore it seems useful to consider nonstatic models which are relatively
simple to analyze but still contain some of the essential features of a
realistic situation.

Accordingly, it is our purpose in this work to present an approach for
modeling the evolution of self--gravitating spheres, which may be regarded
as a ``compromise'' between the two approaches mentioned above (analytical
and numerical).

Indeed, the proposed method, starting from any interior (analytical) static
spherically
symmetric (``seed'') solution to Einstein equations, leads to a system of
ordinary differential equations for quantities evaluated at the boundary
surface of the fluid distribution,
whose solution (numerical), allows for modeling the dynamics of
self--gravitating
spheres, whose static limit is the original ``seed'' solution.

The approach is based on the introduction of a set of conveniently defined
``effective'' variables, which are effective pressure and energy density, and an
heuristic ansatzs on the latter
\cite{barreto}, whose rationale and justification become intelligible
within the context of the post--quasistatic appproximation defined below. In
the quasistatic approximation
(see next section), the effective variables coincide with the corresponding
physical variables (pressure and density) and  therefore the method may be
regarded as an iterative method
with each consecutive step corresponding to a stronger departure from
equilibrium.
In  this work we shall restrain ourselves to the post--quasistatic level
(see next section for details).

At this point it is important to stress a crucial difference between this
method and the one proposed many years ago with a similar structure, but
based on radiative Bondi coordinates
(see \cite{Herrera} and references therein): in the latter the introduced
effective variables do not coincide with the corresponding physical
variables in the quasistatic
approximation (they do coincide in the static limit), and accordingly the
ansatz on those variables remains as an heuristic assumption, only
justified by the eventual suitability of the
obtained models.

The fluid distribution under consideration will be assumed to
be dissipative. Indeed, dissipation due to the emission of massless
particles (photons and/or neutrinos) is a characteristic process in the
evolution of massive stars. In fact, it seems that the only plausible
mechanism to carry away the bulk of the binding energy of the collapsing
star, leading to a neutron star or black hole is neutrino emission
\cite{1}. Consequently, in this paper, the matter distribution forming
the selfgravitating object will be described as a dissipative
fluid.

In the diffusion approximation, it is assumed that the energy flux of
radiation (as that of
thermal conduction) is proportional to the gradient of temperature. This
assumption is in general very sensible, since the mean free path of
particles responsibles for the propagation of energy in stellar
interiors is in general very small as compared with the typical
length of the object.
Thus, for a main sequence star as the sun, the mean free path of
photons at the centre, is of the order of $2\, cm$. Also, the
mean free path of trapped neutrinos in compact cores of densities
about $10^{12} \, g.cm.^{-3}$ becomes smaller than the size of the stellar
core \cite{3,4}.

Furthermore, the observational data collected from supernovae 1987A
indicates that the regime of radiation transport prevailing during the
emission process, is closer to the diffusion approximation than to the
streaming out limit \cite{5}.

However in many other circumstances, the mean free path of particles
transporting energy may be large enough as to justify the  free streaming
approximation. Therefore our formalism
will include simultaneously both limiting  cases of radiative transport
(diffusion and streaming out), allowing for describing a wide range
situations.

Besides the usual physical variables (energy density, pressure, velocity,
heat flow, etc.) we shall also incorporate into discussion other quantities
which are expected to play an
important role in the evolution of evolving self--gravitating systems, such
as the Weyl tensor, the shear of the fluid and the Tolman mass. Therefore
these quantities
will be calculated and used in the process of modeling. It is also worth
mentioning that although the most common method of solving Einstein's
equations is to use commoving coordinates
(e.g.
\cite{May},
\cite{Bonnor}), we shall use noncomoving coordinates, which implies that
the velocity of any fluid element (defined with respect to a conveniently
chosen set of observers) has to be
considered as a relevant physical variable (\cite{Knutsen}).

The plan of the paper is as follows. In Section 2 we define the conventions
and give the field equations and expressions for the kinematical and
physical variables we shall use, in
noncomoving coordinates. The proposed approach is presented and explained
in Section 3. In Section 4 we illustrate the method by means of three
examples. Finally a discussion of results
is presented in Section 5.

\section{Relevant  Equations and Conventions}
\subsection{The field equations}
We consider spherically symmetric distributions of collapsing
fluid, which for sake of completeness we assume to be locally anisotropic,
undergoing dissipation in the form of heat flow and/or free streaming
radiation, bounded by a
spherical surface $\Sigma$.

\noindent
The line element is given in Schwarzschild--like coordinates by

\begin{equation}
ds^2=e^{\nu} dt^2 - e^{\lambda} dr^2 -
r^2 \left( d\theta^2 + sin^2\theta d\phi^2 \right),
\label{metric}
\end{equation}

\noindent
where $\nu(t,r)$ and $\lambda(t,r)$ are functions of their arguments. We
number the coordinates: $x^0=t; \, x^1=r; \, x^2=\theta; \, x^3=\phi$.

\noindent
The metric (\ref{metric}) has to satisfy Einstein field equations

\begin{equation}
G^\nu_\mu=-8\pi T^\nu_\mu,
\label{Efeq}
\end{equation}

\noindent
which in our case read \cite{Bo}:

\begin{equation}
-8\pi T^0_0=-\frac{1}{r^2}+e^{-\lambda}
\left(\frac{1}{r^2}-\frac{\lambda'}{r} \right),
\label{feq00}
\end{equation}

\begin{equation}
-8\pi T^1_1=-\frac{1}{r^2}+e^{-\lambda}
\left(\frac{1}{r^2}+\frac{\nu'}{r}\right),
\label{feq11}
\end{equation}

\begin{eqnarray}
-8\pi T^2_2  =  -  8\pi T^3_3 = & - &\frac{e^{-\nu}}{4}\left(2\ddot\lambda+
\dot\lambda(\dot\lambda-\dot\nu)\right) \nonumber \\
& + & \frac{e^{-\lambda}}{4}
\left(2\nu''+\nu'^2 -
\lambda'\nu' + 2\frac{\nu' - \lambda'}{r}\right),
\label{feq2233}
\end{eqnarray}

\begin{equation}
-8\pi T_{01}=-\frac{\dot\lambda}{r},
\label{feq01}
\end{equation}

\noindent
where dots and primes stand for partial differentiation with respect
to $t$ and $r$,
respectively.

\noindent
In order to give physical significance to the $T^{\mu}_{\nu}$ components
we apply the Bondi approach \cite{Bo}.

\noindent
Thus, following Bondi, let us introduce purely locally Minkowski
coordinates ($\tau, x, y, z$)

$$d\tau=e^{\nu/2}dt\,;\qquad\,dx=e^{\lambda/2}dr\,;\qquad\,
dy=rd\theta\,;\qquad\, dz=rsin\theta d\phi.$$

\noindent
Then, denoting the Minkowski components of the energy tensor by a bar,
we have

$$\bar T^0_0=T^0_0\,;\qquad\,
\bar T^1_1=T^1_1\,;\qquad\,\bar T^2_2=T^2_2\,;\qquad\,
\bar T^3_3=T^3_3\,;\qquad\,\bar T_{01}=e^{-(\nu+\lambda)/2}T_{01}.$$

\noindent
Next, we suppose that when viewed by an observer moving relative to these
coordinates with proper velocity $\omega$ in the radial direction, the physical
content  of space consists of an anisotropic fluid of energy density $\rho$,
radial pressure $P_r$, tangential pressure $P_\bot$,  radial heat flux
$\hat q$ and unpolarized radiation of energy density $\hat\epsilon$
traveling in the radial direction. Thus, when viewed by this moving
observer the covariant tensor in
Minkowski coordinates is

\[ \left(\begin{array}{cccc}
\rho + \hat\epsilon    &  -\hat q - \hat\epsilon  &   0     &   0    \\
-\hat q - \hat\epsilon &  P_r + \hat\epsilon    &   0     &   0    \\
0       &   0       & P_\bot  &   0    \\
0       &   0       &   0     &   P_\bot
\end{array} \right). \]

\noindent
Then a Lorentz transformation readily shows that

\begin{equation}
T^0_0=\bar T^0_0= \frac{\rho + P_r \omega^2 }{1 - \omega^2} +
\frac{2 Q \omega e^{\lambda/2}}{(1 - \omega^2)^{1/2}} + \epsilon,
\label{T00}
\end{equation}

\begin{equation}
T^1_1=\bar T^1_1=-\frac{ P_r + \rho \omega^2}{1 - \omega^2} -
\frac{2 Q \omega e^{\lambda/2}}{(1 - \omega^2)^{1/2}}-\epsilon,
\label{T11}
\end{equation}

\begin{equation}
T^2_2=T^3_3=\bar T^2_2=\bar T^3_3=-P_\bot,
\label{T2233}
\end{equation}

\begin{equation}
T_{01}=e^{(\nu + \lambda)/2} \bar T_{01}=
-\frac{(\rho + P_r) \omega e^{(\nu + \lambda)/2}}{1 - \omega^2} -
\frac{Q e^{\nu/2} e^{\lambda}}{(1 - \omega^2)^{1/2}} (1 + \omega^2)
-e^{(\nu + \lambda)/2} \epsilon,
\label{T01}
\end{equation}
\noindent
with

\begin{equation}
Q \equiv \frac{\hat q e^{-\lambda/2}}{(1 - \omega^2)^{1/2}}
\label{defq}
\end{equation}
and
\begin{equation}
\epsilon\equiv\hat\epsilon\frac{(1+\omega)}{(1-\omega)}.
\label{defepsilon}
\end{equation}

\noindent
Note that the coordinate velocity in the ($t,r,\theta,\phi$) system, $dr/dt$,
is related to $\omega$ by

\begin{equation}
\omega=\frac{dr}{dt}\,e^{(\lambda-\nu)/2}.
\label{omega}
\end{equation}

\noindent
Feeding back (\ref{T00}--\ref{T01}) into (\ref{feq00}--\ref{feq01}), we get
the field equations in  the form

\begin{equation}
\frac{\rho + P_r \omega^2 }{1 - \omega^2} +
\frac{2 Q \omega e^{\lambda/2}}{(1 - \omega^2)^{1/2}} +
\epsilon=-\frac{1}{8 \pi}\Biggl\{-\frac{1}{r^2}+e^{-\lambda}
\left(\frac{1}{r^2}-\frac{\lambda'}{r} \right)\Biggr\},
\label{fieq00}
\end{equation}

\begin{equation}
\frac{ P_r + \rho \omega^2}{1 - \omega^2} +
\frac{2 Q \omega e^{\lambda/2}}{(1 -
\omega^2)^{1/2}}+\epsilon=-\frac{1}{8 \pi}\Biggl\{\frac{1}{r^2} - e^{-\lambda}
\left(\frac{1}{r^2}+\frac{\nu'}{r}\right)\Biggr\},
\label{fieq11}
\end{equation}

\begin{eqnarray}
P_\bot = -\frac{1}{8 \pi}\Biggl\{\frac{e^{-\nu}}{4}\left(2\ddot\lambda+
\dot\lambda(\dot\lambda-\dot\nu)\right) \nonumber \\
 - \frac{e^{-\lambda}}{4}
\left(2\nu''+\nu'^2 -
\lambda'\nu' + 2\frac{\nu' - \lambda'}{r}\right)\Biggr\},
\label{fieq2233}
\end{eqnarray}

\begin{equation}
\frac{(\rho + P_r) \omega e^{(\nu + \lambda)/2}}{1 - \omega^2} +
\frac{Q e^{\nu/2} e^{\lambda}}{(1 - \omega^2)^{1/2}} (1 + \omega^2)
+e^{(\nu + \lambda)/2} \epsilon=-\frac{\dot\lambda}{8 \pi r}.
\label{fieq01}
\end{equation}

Observe that if $\nu$ and $\lambda$ are fully specified, then
(\ref{fieq00}--\ref{fieq01}) becomes a system of  algebraic equations for
the physical variables $\rho$, $P_r$, $P_\bot$,
$\omega$, $Q$ and $\epsilon$. Obviuosly, in the most general case when all
these variables are non--vanishing, the system is underdetermined, and two
equations of state should be
given. In general, whenever $Q\not=0$ a transport equation has to be
assumed. In the case originally considered by  Bondi \cite{Bo} (locally
isotropic fluid and free streaming regime,
$Q=0$) the system is closed. For  the adiabatic ($\epsilon=Q=0$), and
locally isotropic fluid ($P_r=P_\bot$) the system is overdetermined, and a
constraint on the physical variables
appears.

\noindent
At the outside of the fluid distribution, the spacetime is that of Vaidya,
given by

\begin{equation}
ds^2= \left(1-\frac{2M(u)}{{\cal R}}\right) du^2 + 2dud{\cal R} -
{\cal R}^2 \left(d\theta^2 + sin^2\theta d\phi^2 \right),
\label{Vaidya}
\end{equation}

\noindent
where $u$ is a coordinate related to the retarded time, such that
$u=constant$ is (asymptotically) a
null cone open to the future and ${\cal R}$ is a null coordinate ($g_{{\cal
R}{\cal R}}=0$). It should
be remarked, however, that strictly speaking, the radiation can be considered
in radial free streaming only at radial infinity.

\noindent
The two coordinate systems ($t,r,\theta,\phi$) and ($u,{\cal
R},\theta,\phi$) are
related at the boundary surface and outside it by

\begin{equation}
u=t-r-2M\,ln \left(\frac{r}{2M}-1\right),
\label{u}
\end{equation}

\begin{equation}
{\cal R}=r.
\label{radial}
\end{equation}

\noindent
In order to match smoothly the two metrics above on the boundary surface
$r=r_\Sigma(t)$, we first require the continuity of the first fundamental
form across that surface.
Then
\begin{equation}
\left[e^{\nu_{\Sigma}}-e^{\lambda{_\Sigma}} \dot r^2_\Sigma\right]\,dt^2=
\left[1-\frac{2M}{R_\Sigma}+2\frac{dR_\Sigma}{du}\right]\,du^2,
\label{junc}
\end{equation}
where $R=R_\Sigma(u)$ is the equation of the boundary surface in
($u,R,\theta,\phi$)
coordinates.

From (\ref{junc}), using (\ref{omega}), (\ref{u}) and (\ref{Vaidya}) it follows
\begin{equation}
e^{\nu_\Sigma}=1-\frac{2M}{R_\Sigma},
\label{enusigma}
\end{equation}
\begin{equation}
e^{-\lambda_\Sigma}=1-\frac{2M}{R_\Sigma}.
\label{elambdasigma}
\end{equation}
Where, from now on, subscript $\Sigma$ indicates that the quantity is
evaluated at the boundary surface $\Sigma$.

Next, the unit vector $n_\mu$, normal to the boundary surface, has components
\begin{equation}
n_\mu^{(+)}=\left(-\beta\,\frac{dR_\Sigma}{du},\, \beta,\, 0,\, 0\right),
\label{nmu+}
\end{equation}
where $+$ indicates that the components are evaluated from the outside of
$\Sigma$, and $\beta$ is given by
\begin{equation}
\beta=\left(1-\frac{2M(u)}{R_\Sigma}+2\frac{dR_\Sigma}{du}\right)^{
-1/2}.
\label{beta}
\end{equation}
The unit vector normal to $\Sigma$, evaluated from the inside, is given by
\begin{equation}
n_\mu^{(-)}=\left(-\dot{r}_{\Sigma}\gamma,\, \gamma,\, 0,\, 0\right),
\label{nmu-}
\end{equation}
with
\begin{equation}
\gamma=\left(e^{-\lambda_\Sigma}-\dot{r}^2_\Sigma\,
e^{-\nu_\Sigma}\right)^{-1/2}.
\label{gamma}
\end{equation}
Let us now define a time--like vector $v^\mu$ such that
\begin{equation}
v^{\mu(+)}=\beta\delta^\mu_u+\beta\,\frac{dR_\Sigma}{du}\,\delta^\mu_R
\label{vmu+}
\end{equation}
and
\begin{equation}
v^{\mu(-)}=\frac{e^{-\nu_\Sigma/2}}{\left(1-\omega^2_\Sigma\right)^{1/2}}
\delta^\mu_t +\frac{\omega_\Sigma e^{-\lambda_\Sigma/2}}
{\left(1-\omega^2_\Sigma\right)
^{1/2}}\delta^\mu_r.
\label{vmu-}
\end{equation}
Then, junction conditions across $\Sigma$, require (besides (\ref{junc}))
\begin{equation}
\left(T_{\mu\nu}n^\mu n^\nu\right)^{(+)}_\Sigma =
\left(T_{\mu\nu}n^\mu n^\nu\right)^{(-)}_\Sigma,
\label{contTn}
\end{equation}
\begin{equation}
\left(T_{\mu\nu}n^\mu v^\nu\right)^{(+)}_\Sigma =
\left(T_{\mu\nu}n^\mu v^\nu\right)^{(-)}_\Sigma,
\label{contTv}
\end{equation}
where the expressions for the energy momentum tensor at both sides of
the boundary surface are
\begin{equation}
T_{\mu\nu}^{(-)} = \left(\rho+P_\bot\right)u_\mu u_\nu - P_\bot g_{\mu\nu} +
\left(P_r-P_\bot\right)s_\mu s_\nu + q_\mu u_\nu + q_\nu u_\mu + \epsilon
l_\nu l_\mu
\label{T-}
\end{equation}
and
\begin{equation}
T_{\mu\nu}^{(+)} = -\frac{1}{4\pi
R^2}\,\frac{dM}{du}\,\delta_\mu^0\delta_\nu^0,
\label{T+}
\end{equation}
with
\begin{equation}
u^\mu=\left(\frac{e^{-\nu/2}}{\left(1-\omega^2\right)^{1/2}},\,
\frac{\omega\, e^{-\lambda/2}}{\left(1-\omega^2\right)^{1/2}},\,0,\,0\right),
\label{umu}
\end{equation}
\begin{equation}
s^\mu=\left(\frac{\omega \, e^{-\nu/2}}{\left(1-\omega^2\right)^{1/2}},\,
\frac{e^{-\lambda/2}}{\left(1-\omega^2\right)^{1/2}},\,0,\,0\right),
\label{smu}
\end{equation}
\begin{equation}
l^\mu=\left(e^{-\nu/2},\,e^{-\lambda/2},\,0,\,0\right),
\label{null}
\end{equation}
where $u^\mu$ denotes the four velocity of the fluid,  $s^\mu$ is a
radially directed space--like vector orthogonal to $u^\mu$, $l^\mu$ is a
null outgoing vector, and
\begin{equation}
q^\mu=Q\,\left(\omega\,e^{(\lambda-\nu)/2},\,1,\,0,\,0\right).
\label{qmu}
\end{equation}
Then it follows from (\ref{contTn}) and (\ref{contTv})
\begin{equation}
[P_r + \hat\epsilon]_\Sigma=
-\left[\frac{1}{4\pi R^2}\,\frac{dM}{du}\,\beta^2\right]_\Sigma,
\label{Psup}
\end{equation}
\begin{equation}
[Q e^{\lambda/2} (1-\omega^2)^{1/2} +\hat\epsilon]_\Sigma=
-\left[\frac{1}{4\pi R^2}\,
\frac{dM}{du}\,\beta^2\right]_\Sigma.
\label{Qsup}
\end{equation}
Eqs. (\ref{junc}), (\ref{Psup}) and (\ref{Qsup}) are the necessary and
sufficient conditions for a smooth matching of the two metrics (\ref{metric})
and (\ref{Vaidya}) on $\Sigma$.
Combining (\ref{Psup}) and (\ref{Qsup}) we get
\begin{equation}
\left[P_r\right]_\Sigma=\left[Q\,e^{\lambda/2}\left(1-\omega^2\right)^
{1/2}\right]_\Sigma,
\label{PQ}
\end{equation}
expressing the discontinuity of the radial pressure in the presence
of heat flow, which is a well known result \cite{Sa}.

\noindent
Next, it will be useful to calculate the radial component of the
conservation law

\begin{equation}
T^\mu_{\nu;\mu}=0.
\label{dTmn}
\end{equation}

\noindent
After tedious but simple calculations we get

\begin{equation}
\left(-8\pi T^1_1\right)'=\frac{16\pi}{r} \left(T^1_1-T^2_2\right)
+ 4\pi \nu' \left(T^1_1-T^0_0\right) +
\frac{e^{-\nu}}{r} \left(\ddot\lambda + \frac{\dot\lambda^2}{2}
- \frac{\dot\lambda \dot\nu}{2}\right),
\label{T1p}
\end{equation}

\noindent
which in the static case becomes

\begin{equation}
P'_r=-\frac{\nu'}{2}\left(\rho+P_r\right)+
\frac{2\left(P_\bot-P_r\right)}{r},
\label{Prp}
\end{equation}

\noindent
representing the generalization of the Tolman--Oppenheimer--Volkof equation
for anisotropic fluids \cite{BoLi}.
\subsection{The kinematical variables}
The components of the shear tensor are defined by

\begin{equation}
\sigma_{\mu\nu}=u_{\mu;\nu}+u_{\nu;\mu}-u_{\mu}a_{\nu}-u_{\nu}a_{\mu}-
\frac{2}{3}{\Theta}P_{\mu\nu},
\label{shear}
\end{equation}

\noindent
where

\begin{equation}
P_{\mu\nu}=g_{\mu\nu}-u{_\mu}u_{\nu}\,;\qquad\,
\Theta=u^{\mu}_{;\mu}\,;\qquad\,a_{\mu}=u^{\nu}u_{\mu;\nu},
\label{P,Theta,a}
\end{equation}

\noindent
denote the projector onto the three space orthogonal to $u^\mu$, the
expansion  and the four acceleration, respectively.

\noindent
A simple calculation gives

\begin{equation}
\Theta=\frac{e^{-\nu/2}}{2\left(1-\omega^2\right)^{1/2}}
\left(\dot\lambda + \frac{2\omega\dot\omega}{1-\omega^2}\right) +
\frac{e^{-\lambda/2}}{2\left(1-\omega^2\right)^{1/2}}
\left(\omega\nu' + 2\omega' +
\frac{2\omega^2\omega'}{1-\omega^2} +
\frac{4\omega}{r}\right),
\label{Theta}
\end{equation}

\begin{equation}
\sigma_{11}= -\frac{2}{3\left(1-\omega^2\right)^{3/2}}
\left[e^{\lambda}e^{-\nu/2}\left(\dot\lambda +
\frac{2\omega\dot\omega}{1-\omega^2}\right) + e^{\lambda/2}
\left(\omega\nu' + \frac{2\omega'}{1-\omega^2}
- \frac{2\omega}{r}\right)  \right],
\label{shear11}
\end{equation}

\begin{equation}
\sigma_{22}=-\frac{e^{-\lambda} r^2 \left(1-\omega^2\right)}{2}
\sigma_{11},
\label{shear22}
\end{equation}

\begin{equation}
\sigma_{33}=-\frac{e^{-\lambda} r^2 \left(1-\omega^2\right)}{2}
\sin^2{\theta} \sigma_{11},
\label{shear33}
\end{equation}

\begin{equation}
\sigma_{00}=\omega^2 e^{-\lambda} e^\nu \sigma_{11},
\label{shear00}
\end{equation}

\begin{equation}
\sigma_{01}=-\omega e^{\left(\nu-\lambda\right)/2} \sigma_{11},
\label{shear01}
\end{equation}

\begin{equation}
a_0=\frac{1}{1-\omega^2}\left[\left(\frac{\omega\dot\omega}{1-\omega^2} +
\frac{\omega^2 \dot\lambda}{2}\right) +
e^{\nu/2} e^{-\lambda/2}
\left(\frac{\omega \nu'}{2} +
\frac{\omega^2 \omega'}{1-\omega^2}\right)
\right],
\label{a0}
\end{equation}

\begin{equation}
a_1=-\frac{1}{1-\omega^2}\left[\left(\frac{\omega \omega'}{1-\omega^2} +
\frac{\nu'}{2}\right) +
e^{-\nu/2} e^{\lambda/2}
\left(\frac{\omega \dot\lambda}{2} +
\frac{\dot\omega}{1-\omega^2}\right)
\right],
\label{a1}
\end{equation}
and for the shear scalar $\sigma$
\be
\sigma =\sqrt{3}\left( \frac \Theta 3-\frac{e^{-\lambda /2}}r\frac
\omega {\sqrt{1-\omega ^2}}\right). \label{eq:esf}
\ee
\subsection{The Tolman mass}
The Tolman mass for a spherically symmetric distribution
of matter is given by (eq.(24) in \cite{To})

\begin{eqnarray}
m_T = & &  4 \pi \int^{r_\Sigma}_{0}{r^2 e^{(\nu+\lambda)/2}
\left(T^0_0 - T^1_1 - 2 T^2_2\right) dr}\nonumber \\
& + & \frac{1}{2} \int^{r_\Sigma}_{0}{r^2 e^{(\nu+\lambda)/2}
\frac{\partial}{\partial t}
\left(\frac{\partial \pounds}{\partial \left[\partial
\left(g^{\alpha \beta} \sqrt{-g}\right) / \partial t\right]}\right)
g^{\alpha \beta}dr},
\label{Tol}
\end{eqnarray}

\noindent
where $\pounds$ denotes the usual gravitational lagrangian density
(eq.(10) in \cite{To}). Although Tolman's formula was introduced
as a measure of the total energy of the system, with no commitment
to its localization, we shall define the mass within a sphere of
radius $r$, completely inside $\Sigma$, as

\begin{eqnarray}
m_T = & &  4 \pi \int^{r}_{0}{r^2 e^{(\nu+\lambda)/2}
\left(T^0_0 - T^1_1 - 2 T^2_2\right) dr}\nonumber \\
& + & \frac{1}{2} \int^{r}_{0}{r^2 e^{(\nu+\lambda)/2}
\frac{\partial}{\partial t}
\left(\frac{\partial \pounds}{\partial \left[\partial
\left(g^{\alpha \beta} \sqrt{-g}\right) / \partial t\right]}\right)
g^{\alpha \beta}dr}.
\label{Tolin}
\end{eqnarray}

\noindent
This extension of the global concept of energy to a local level
\cite{Coo} is suggested by the conspicuous role played by
$m_T$ as the ``effective gravitational mass'', which will be
exhibited below.
 Even though Tolman's definition is not
without its problems \cite{Coo,Deu}, we shall see that $m_T$,
as defined by (\ref{Tolin}), is a good measure of the
active gravitational mass, at least for the systems under
consideration.

\noindent
Let us now evaluate expression (\ref{Tolin}). The first
integral in that expression

\begin{equation}
I \equiv 4 \pi \int^{r}_{0}{r^2 e^{(\nu+\lambda)/2}
\left(T^0_0 - T^1_1 - 2 T^2_2\right) dr},
\label{I}
\end{equation}

\noindent
may be transformed to give (see \cite{inhomo} for details)

\begin{eqnarray}
I = & & e^{(\nu+\lambda)/2} \left[m(r,t) - \frac{4 \pi}{3} r^3 T^1_1\right]
\nonumber \\
& - & \int^r_0{e^{(\lambda-\nu)/2} \frac{r^2}{2}
\left(\ddot\lambda + \frac{\dot\lambda^2}{2}
- \frac{\dot\lambda \dot\nu}{2}\right) dr},
\label{Ifin}
\end{eqnarray}
where the mass function $m$, as usually is defined by

\begin{equation}
e^{-\lambda(r,t)}= 1-2 m(r,t)/r.
\label{mass}
\end{equation}

Next,
\noindent
 from
(eq.(13) in \cite{To})

\begin{equation}
\frac{\partial}{\partial t}
\left(\frac{\partial \pounds}{\partial \left[\partial
\left(g^{\alpha \beta} \sqrt{-g}\right) / \partial t\right]}\right) =
- \Gamma^{0}_{\alpha \beta} + \frac{1}{2} \delta^0_\alpha
\Gamma^{\sigma}_{\beta \sigma}
+ \frac{1}{2} \delta^0_\beta
\Gamma^{\sigma}_{\alpha \sigma},
\label{13To}
\end{equation}

\noindent
and so the second integral $(II)$ in (\ref{Tolin}) may be expressed as

\begin{equation}
II = \frac{1}{2} \int^r_0{r^2 e^{(\lambda-\nu)/2}
\left(\ddot\lambda + \frac{\dot\lambda^2}{2} -
\frac{\dot\lambda \dot\nu}{2}\right)  dr}.
\label{II}
\end{equation}

\noindent
Thus

\begin{equation}
m_T \equiv I + II = e^{(\nu + \lambda)/2}
\left[m(r,t) - 4 \pi r^3 T^1_1\right].
\label{I+II}
\end{equation}

\noindent
This is, formally, the same
expression for $m_T$ in terms of $m$ and $T^1_1$ that
appears in the static (or quasi-static) case
(eq.(25) in \cite{HeSa}).

\noindent
Replacing $T^1_1$ by (\ref{feq11}) and $m$ by (\ref{mass}),
one  also finds

\begin{equation}
m_T = e^{(\nu - \lambda)/2} \, \nu' \, \frac{r^2}{2}.
\label{mT}
\end{equation}

\noindent
This last equation brings out the physical meaning of $m_T$ as the
active gravitational mass. Indeed, it can be easily shown \cite{Gro}
that the gravitational acceleration $a$ of a test particle,
instantaneously at rest in a static gravitational field, as measured
with standard rods and coordinate clock is given by

\begin{equation}
a = - \frac{e^{(\nu - \lambda)/2} \, \nu'}{2} = - \frac{m_T}{r^2}.
\label{a}
\end{equation}

\noindent
A similar conclusion can be obtained by inspection of Eq. (\ref{Prp})
(valid only in the static or quasi-static case) \cite{Lig}.
In fact, the first term on the right side of this equation
(the ``gravitational force'' term) is a product of the ``passive''
gravitational mass density $(\rho + P_r)$ and a term proportional
to $m_T/r^2$.
\subsection{The Weyl tensor}
Since the publication of Penrose`s work \cite{Penrose}, there has been an
increasing interest in studying the possible role of Weyl tensor (or some
function of it)
in the evolution of self-gravitating systems \cite{arrow}. This interest is
reinforced by the fact that for spherically symmetric distribution of
fluid, the Weyl tensor may be defined
 exclusively in terms of the density contrast and the local anisotropy  of
the pressure (see below), which in turn are known to affect the fate of
gravitational collapse \cite{large}.

Now, using Maple V, it is found
 that all non--vanishing components of the Weyl tensor are
proportional to

\begin{eqnarray}
W \equiv \frac{r}{2} C^{3}_{232} & = & W_{(s)} + \frac{r^3 e^{-\nu}}{12}
\left(\ddot\lambda + \frac{\dot\lambda^2}{2} -
\frac{\dot\lambda \dot\nu}{2}\right)
\label{W}
\end{eqnarray}
\noindent
where

\begin{equation}
W_{(s)} =
\frac{r^3 e^{-\lambda}}{6}
\left( \frac{e^\lambda}{r^2} - \frac{1}{r^2} +
\frac{\nu' \lambda'}{4} - \frac{\nu'^2}{4} -
\frac{\nu''}{2} - \frac{\lambda'}{2r} + \frac{\nu'}{2r} \right),
\label{Ws}
\end{equation}

\noindent
 corresponds
to the contribution in the static case.

Also, from the field equations and the definition of the Weyl tensor it can
be easily shown that (see \cite{inhomo} for details)

\begin{equation}
W = - \frac{4 \pi}{3} \int^r_0{r^3 \left(T^0_0\right)' dr} +
\frac{4 \pi}{3} r^3 \left(T^2_2 - T^1_1\right).
\label{Wint}
\end{equation}

\section{The method}

We have now available all the ingredients required to present our method,
however before doing so some general considerations will be necessary.

\subsection{Equilibrium and departures from equilibrium}
The simplest situation, when dealing with self--gravitating spheres, is that
of equilibrium (static case). In our notation that means that
$\omega=\epsilon=Q=0$,  all time derivatives
vanishes, and we obtain the generalized Tolman--Oppenheimer--Volkof equation
(\ref{Prp}).

Next, we have the quasistatic regime. By this we mean that the sphere
changes slowly, on a time scale that is very long compared to the typical
time in which the sphere reacts to a slight perturbation of hydrostatic
equilibrium, this typical time scale is called hydrostatic time scale
\cite{Weigert} (sometimes this time scale is also referred to as dynamical
time scale, e.g. see the third
reference in
\cite{Weigert}). Thus, in this regime  the system is always very close to
hydrostatic
equilibrium and its evolution may be regarded as a sequence of static
models linked by (\ref{fieq01}). This assumption is very sensible because
the hydrostatic
time scale is very small for many phases of the life of the star.
It is of the order of $27$ minutes for the Sun, $4.5$ seconds for a white dwarf
and $10^{-4}$ seconds for a neutron star of one solar mass and $10$ Km radius.
It is well known that any of the stellar configurations mentioned above,
generally,
change on a time scale that is very long compared to their respective
hydrostatic time scales. Let us now translate this assumption in conditions
to $\omega$ and metric functions.

First of all, slow contraction means that the radial velocity $\omega$ as
measured by the Minkowski observer is always much smaller than the velocity
of light ($\omega \ll 1$). Therefore we have to neglect terms of the order
$O(\omega^2)$.

Then (\ref{T1p}) yields
\begin{equation}
\ddot\lambda + \frac{\dot\lambda^2}{2} -
\frac{\dot\nu \dot\lambda}{2} = 8 \pi r e^{\nu}
\left[(P_r +\epsilon)'+ \left(\rho + P_r + 2 \epsilon\right) \frac{\nu'}{2} -
2 \frac{P_\bot - P_r - \epsilon}{r}\right].
\label{lsps}
\end{equation}
(observe the contribution of $\epsilon$ to, both,
$P_r$ and $\rho$ and the fact that $\epsilon$, $\omega$ and $Q$ are of the
same order of smallness, in this approximation).

Since by assumption, in this regime the system is always (not only at a
given time $t$)
in equilibrium (or very close to), (\ref{Prp}) and (\ref{lsps}) imply then,
for an arbitrary slowly evolving configuration
\begin{equation}
\ddot\lambda \approx \dot\nu \dot\lambda \approx
\dot\lambda^2 \approx 0,
\label{lp0}
\end{equation}
and of course, time derivatives of any order of the left hand side of the
hydrostatic equilibrium equation must also vanish, for otherwise the
system will deviate from equilibrium. This condition implies, in particular,
that we must demand in this regime
$$
\ddot\nu \approx 0.
$$

Finally, from the time derivative of (\ref{feq01}), and using (\ref{T01}),
it follows that

\begin{equation}
\dot\omega \approx O(\ddot\lambda, \dot\lambda \omega, \dot\nu \omega).
\label{Omo}
\end{equation}
which implies that we have also to neglect terms linear in the acceleration.
On purely physical considerations, it is obvious that the vanishing of
$\dot\omega$ is required to keep the system always in equilibrium.

Thus, in the quasistatic regime we have to assume
\begin{equation}
O(\omega^2) = {\dot\lambda}^2 = {\dot\nu}^2 =
\dot\lambda \dot\nu = \ddot\lambda = \ddot\nu = 0,
\label{om2}
\end{equation}
implying that the system remains in (or very close to) equilibrium.
However, during their evolution, self--gravitating objects may pass
through phases of intense dynamical activity, with time scales of the order
of magnitude of (or even smaller than) the hydrostatic time scale, and  for
which the quasi--static
approximation is clearly not reliable (e.g.,the collapse of very massive
stars \cite{Iben} and the quick collapse phase
preceding neutron star formation, see for example \cite{myra} and
references therein). In these cases it is mandatory to take into account
terms which describe departure from
equilibrium.

\subsection{The effective variables and the post--quasistatic approximation}

Let us now define the following effective variables:

\begin{equation}
\tilde\rho=T^0_0= \frac{\rho + P_r \omega^2 }{1 - \omega^2} +
\frac{2 Q \omega e^{\lambda/2}}{(1 - \omega^2)^{1/2}} + \epsilon,
\label{rhoeffec}
\end{equation}

\begin{equation}
\tilde P=-T^1_1=\frac{ P_r + \rho \omega^2}{1 - \omega^2} +
\frac{2 Q \omega e^{\lambda/2}}{(1 - \omega^2)^{1/2}}+\epsilon.
\label{peffec}
\end{equation}

 In the quasistatic regime the effective variables satisfy the same
equation (\ref{Prp}) as the corresponding physical variables (taking into
account the contribution of $\epsilon$ to the ``total'' energy density and
radial pressure, whenever
the free streaming approximation is being used). Therefore in
the quasistatic situation (and obviously
 in  the static too), effective and physical variables share the same
radial dependence.
Next, feeding back (\ref{rhoeffec}) and (\ref{peffec}) into (\ref{fieq00})
and (\ref{fieq11}), these two equations may be formally integrated , to
obtain:

\begin{equation}
m = 4 \pi \int^{r}_{0}{r^2 \tilde\rho dr}, \\
\label{m}
\end{equation}

\begin{equation}
\nu = \nu_\Sigma + \int^{r}_{r_\Sigma}\frac{2 (4 \pi r^3  \tilde
P+m)}{r(r-2m)} dr.
\label{nu}
\end{equation}
From where it is obvious that for a given radial dependence of the
effective variables, the radial dependence of metric functions become
completely determined.

With this last comment in mind, we shall define the post--quasistatic regime
as that corresponding to  a system out of equilibrium (or quasiequilibrium)
but whose effective variables
share the same radial dependence as the corrresponding physical variables
in the state of equilibrium (or quasiequilibrium). Alternatively it may be
said that the system in the
post--quasistatic regime is characterized by metric functions whose radial
dependence is the same as the metric functions corresponding to the static
(quasistatic) regime. The rationale
behind this definition is not difficult to grasp: we look for a regime
which although out of equilibrium, represents the closest possible
situation to a quasistatic evolution (see more
on this point in the last Section).

\subsection{The algorithm}

Let us now outline the approach that we propose:
\begin{enumerate}
\item  Take an interior solution to Einstein equations, representing a fluid
distribution of matter in equilibrium, with a given

$$\rho_{st}=\rho(r)\,\qquad\, P_{r\, st}= P_{r}(r)$$

\item  Assume that the $r$ dependence of $\tilde P$ and $\tilde\rho$ is the
same as that of $P_{r\, st}$ and $\rho_{st}$, respectively.

\item  Using equations (\ref{nu}) and (\ref{m}), with the $r$ dependence of
$\tilde P$ and $\tilde\rho$, one gets $m$ and $\nu$ up to some functions of
$t$, which will be specified below.

\item  For these functions of $t$ one has three ordinary differential equations
(hereafter referred to as surface equations), namely:
\begin{enumerate}
\item  Equation (\ref{omega}) evaluated  on $r=r_{\Sigma}$.

\item  Equation (\ref{T1p}) evaluated on $r=r_{\Sigma}$.

\item The equation relating the total mass loss rate with the energy flux
through the boundary surface.
\end{enumerate}
\item Depending on the kind of matter under consideration, the system of
surface equations described above may be closed with the additional
information provided by the transport equation
and/or the equation of state for the anisotropic pressure and/or
additional information about some of the physical variables evaluated on
the boundary surface (e.g. the
luminosity).

\item Once the system of surface equations is closed, it may be integrated for
any particular initial data.

\item  Feeding back the result of integration in the expressions for $m$ and
$\nu$, these two functions are completely determined.

\item  With the input from the point 7 above, and using field equations,
together with the equations of state and/or transport equation, all
physical variables may be found for any piece of
matter distribution.
\end{enumerate}
\subsection{The Surface equations}

 As it should be clear from the above the crucial point in the algorithm is
the system of surface equations. So, let us specify them now.

Introducing the dimenssionles variables

$$A=r_{\Sigma}/m_{\Sigma}(0),$$
$$F=1-2M/A,$$
$$M=m_{\Sigma}/m_{\Sigma}(0),$$
$$\Omega=\omega_{\Sigma},$$
$$\alpha=t/m_{\Sigma}(0),$$
where  $m_{\Sigma}(0)$ denotes the total initial mass, we obtain  the first
surface equation by
evaluating (\ref{omega}) at $r=r_{\Sigma}$, one gets

\begin{equation}
\frac{dA}{d\alpha}=F\Omega. \label{eq:first}
\end{equation}

Next, using junction conditions, one obtains from (\ref{mass}),
(\ref{fieq00}) and (\ref{fieq01})
evaluating at $r=r_{\Sigma}$, that

\be
\frac{dM}{d\alpha}=-F (1+\Omega)\hat E,
\label{flux}
\ee
with
\be
\hat E=4 \pi r^{2}_{\Sigma} (\hat \epsilon_{\Sigma}+ \hat q_{\Sigma}),
\ee
where the first and second term on the right of (\ref{flux}), represent the
gravitational redshift and the Doppler shift corrections, respectively.

Then, defining the luminosity perceived by an observer at infinity as
$$L=-\frac{dM}{d\alpha}.$$
we obtain the second surface equation in the form

\be
\frac{dF}{d\alpha}=\frac FA(1-F)\Omega +2 L/A. \label{eq:second}
\ee

The third surface equation may be obtained by evaluating at the boundary
surface, the conservation law
$T_{1;\mu }^\mu=0$, which reads
\begin{equation}
\tilde P^{'} + \frac{(\tilde\rho + \tilde P)(4\pi r^3\tilde P + m)}{r(r-2m)}
=\frac{e^{-\nu}}{4\pi r(r-2m)}\left( \ddot m +\frac{3\dot m^2}{r-2m}-
\frac{\dot m \dot \nu}{2}\right) + \frac{2}{r}(P_\bot-\tilde P). \label{eq:TOV}
\end{equation}

Now, in the following section we consider two relatively simple models with
a separable effective density, i.e., $\tilde\rho=f(t)h(r)$; thus
equation (\ref{eq:TOV}) evaluated at the boundary surface leads to
\begin{equation}
\frac{d\Omega}{d\alpha}=\Omega^2\left[\frac{8F}{A}+2Fk(r_{\Sigma})+4\pi\tilde
\rho_{\Sigma} A(3-\Omega^2)\right]
-\frac{F}{\tilde\rho_{\Sigma}}\left[R-\frac{2}{A}\left(P_{\bot
\Sigma}-\tilde\rho_{\Sigma} \Omega^2
-\frac{\bar E(1+\Omega)}{4\pi r_{\Sigma}^2}\right)\right],
\label{eq:TOV_a}
\end{equation}
where
\begin{equation}
R=\left[\tilde P^{'} + \frac{\tilde P+\tilde\rho}{1-2m/r}(4\pi r\tilde P
+\frac{m}{r^2})\right]_{\Sigma},
\label{R}
\end{equation}

\be
\bar E=\hat E (1+\Omega),
\ee
and
\begin{equation}
k(r_{\Sigma})=\frac{d
}{dr_{\Sigma}}ln\left(\frac{1}{r_{\Sigma}}\int^{r_{\Sigma}}_0 dr r^2
h(r)/h(r_{\Sigma})\right).
\end{equation}

Before analyzing specific models, some interesting conclusions can be
obtained at this level of
generality.
One of these conclusions concerns the condition of bouncing
at the surface which, of course, is related to the occurrence of a minimum
radius $A$. According to (\ref{eq:first}) this requires $\Omega=0$, and we have
\begin{equation}
\frac{d^{2} A}{d{\alpha^2}} = F\frac{d \Omega}{d \alpha},
\end{equation}
or using (\ref{eq:TOV_a})
\begin{equation}
\frac{d\Omega}{d\alpha}(\Omega=0)=-\frac{F}{\tilde\rho_{\Sigma}}\left[R-\frac
{2}{A}\left(P_{\bot \Sigma}
-\frac{\hat E}{4\pi {r_{\Sigma}}^2}\right)\right]. \label{eq:TOV_a0}
\end{equation}
Observe that a positive  energy flux ($\hat E$)
tends to decrease the radius of the
sphere, i.e., it favors the compactification of the object, which is easily
understandable. The same happens
when $R>0$ or $P_{\bot \Sigma}<0$. The opposite effect occurs when these
quantities have the opposite
signs. Now, for a positive energy flux the sphere
can only bounce at its surface when
$$\frac{d\Omega}{d\alpha}(\Omega=0) \ge 0.$$
According to (\ref{eq:TOV_a0})
this is
equivalent to
\begin{equation}
-R(\Omega=0)+ \frac{2P_{\bot \Sigma}}{A}\ge 0.
\label{bounce}
\end{equation}

A physical meaning can be associated to this equation as follows. For
non--radiating, static configuration, $R$ as defined by (\ref{R}) consists
of two parts.
The first term which together with $-[2(P_{\bot}-P_{r})/r]_\Sigma$
represents the hydrodynamical force (see (\ref{Prp})) and the second which
is of course the gravitational force. The
resulting force in the sense of increasing $r$ is precisely $-R+
[2(P_{\bot}-P_{r})/r]_\Sigma$, if this is positive a net outward
acceleration occurs and vice--versa.
Equation (\ref{bounce}) is the natural generalization of this result for
general non--static configurations.

As mentioned before, besides the surface equations, in some cases
(depending on the type of matter under consideration) further information
has to be provided in the form of equation
of state for the tangential stresses and/or transport equation. In the next
Section we shall illustrate our method with three examples, one of which
refers to an anisotropic fluid, and
for which we shall further assume the equation of state (see
\cite{b93},\cite{chew82})

\be
P_\bot - P_r = \frac{C (\tilde{P} + \tilde{\rho})(4 \pi r^{3} \tilde{P} +
m)}{(r-2m)},
\label{eq:EDE}
\ee
where $C$ is a constant.

\section{Examples}

The only purpose of the present Section is to illustrate the proposed
method. For simplification we shall consider only the adiabatic case
($\epsilon=Q=0$). For all these
models we shall calculate the physical and geometrical variables  for any
piece of matter, as function of the timelike coordinate. In spite of the
simplicity of the models, some
interesting conclusions about the physical meaning of different variables
may be reached.

One of the models has as the ``seed'' solution  the well known Schwarzshild
interior solution, whose properties have been extensively discussed in the
literature. The second example is
based on an anisotropic fluid without radial pressure. Models of this kind
have also been discussed extensively since the original Einstein paper (see
\cite{Einstein}). Finally, the
third example represents the dynamic version of the Tolman VI static
solution (\cite{Tolman}), whose equation of state, as is well known,
approaches that for highly compressed Fermi
gas.

\subsection{Schwarzschild--type model}

This model is inspired in the well known interior Schwarzschild solution.
Accordingly we take
\begin{equation}
\tilde \rho = f(t),
\end{equation}
where $f$ is an arbitrary function of $t$.
And the expression for $\tilde P$ is
\begin{equation}
\frac{\tilde P + \frac{1}{3}\tilde\rho}{\tilde P + \tilde\rho}=
\left(1-\frac{8\pi}{3}\tilde\rho r^2\right)^{1/2}k(t), \label{eq:prep}
\end{equation}
where  $k$ is a function of $t$  to be determined from junction conditions
(\ref{PQ}), which in terms of effective variables, becomes
\begin{equation}
\tilde P_{\Sigma}=\tilde\rho_{\Sigma}\Omega^2.
 \label{eq:boun}
\end{equation}

Thus, using (\ref{eq:prep}) and (\ref{eq:boun}) we have for the effective
variables
\begin{equation}
\tilde\rho=\frac{3(1-F)}{8\pi r_{\Sigma}^2},
\end{equation}
\begin{equation}
\tilde P=\frac{\tilde\rho}{3}\Biggl\{\frac{\chi {F}^{1/2} -3\psi\xi}
{\psi\xi -\chi {F}^{1/2}}\Biggr\}, \label{eq:effepre}
\end{equation}
with
$$
\xi=[1-(1-F)(r/r_{\Sigma})^2]^{1/2}
$$
and
$$\chi=3(\Omega^2+1)(1-F),$$

$$\psi=(3\Omega^2+1)(1-F).$$

And for the metric functions $m$ and $\nu$ we get, using (\ref{m}) and
(\ref{nu})
\begin{equation}
m=m_{\Sigma}(r/r_{\Sigma})^3, \label{eq:mass_sch}
\end{equation}
\begin{equation}
e^{\nu}=\Biggl\{\frac{\chi {F}^{1/2}-\psi\xi}{2(1-F)}
\Biggr\}^{2}. \label{eq:nu_sch}
\end{equation}

The third surface equation for this model becomes
\be
\frac{d\Omega}{d\alpha}=\frac{\Omega^2}{2A}(7-3\Omega^2+3F(\Omega^2-1)).
\ee

This equation  together with (\ref{eq:first}) and (\ref{eq:second}) form
the set of surface equations for this model.
We have integrated it numerically and from this integration, all physical
variables are found for any piece of the fluid distribution, following the
algorithm described above.

Figures (1)--(5) exhibit the behaviour of $\rho$, $P$, $\omega$, $\sigma$
and $W$ for an initially contracting configuration, as function of $\alpha$
and different pieces of matter.

Figure (6) shows the profile of $\omega$ as function of $r/r_{\Sigma}$ for
$\alpha=10$.

\subsection{Lemaitre--Florides--type model}

This model  has as the ``seed'' solution a configuration with homogeneous
energy density and vanishing radial pressure. Configurations of this kind
were suggested by the first time by
Lemaitre (\cite{Lemaitre}).

The corresponding effective variables now are:
\begin{equation}
\tilde \rho = f(t)
\label{lrho}
\end{equation}

and

\begin{equation}
\tilde P = 0.
\label{lp}
\end{equation}

Observe that in this case, because of (\ref{lrho}) and (\ref{lp}), it
follows from (\ref{rhoeffec}) and (\ref{peffec}) that the radial pressure
is discontinuos at the boundary
surface, with
\begin{equation}
P_{r \Sigma}=-3 (1- F) \Omega^2/8 \pi r_{\Sigma}^2,
\label{sur.tension}
\end{equation}
for otherwise either $\rho_{\Sigma}$ or $\Omega$ should vanish at
$\Sigma$. Therefore the only way to ``dynamize'' this model is by relaxing
boundary conditions, allowing for the
presence of  a kind of surface tension.

Once the effective variables are defined, we only need  the value of the
tangential pressure at the boundary to close the system of surface
equations. This is obtained by evaluating
(\ref{eq:EDE}) at $\Sigma$.

Next, following the algorithm, all physical variables may be found for any
piece of material as functions of the timelike coordinate. Although we are
not going to exhibit them here,
because the graphics are not particularly illuminating, we wanted to
present an example which, besides the fact that implies an anisotropic
fluid, requires the introduction of a surface
tension, for allowing  the application of the algorithm.

\subsection{Tolman VI--type model}

Our last example is based on the Tolman VI solution. Accordingly the
effective variables for this model will be

\begin{equation}
\tilde\rho=\frac{3g(t)}{r^2}
\label{tolmanrho}
\end{equation}

and

\begin{equation}
\tilde P = \frac{g(9- b K (r/r_{\Sigma}))}{(9-b (r/r_{\Sigma}))  r^2},
\end{equation}
where $g$  and $b$ are functions of $\alpha$, to be obtained from
(\ref{eq:boun}). Then,
\begin{equation}
\tilde\rho=\frac{3(1-F)}{24\pi r^2}.
\end{equation}
Using  (\ref{m}) and (\ref{nu}) we get
\begin{equation}
m=m_{\Sigma} r/r_{\Sigma},
\end{equation}

\begin{equation}
{\nu}= ln F + \frac{8 \pi g}{F} \Biggl\{4 ln(r/r_{\Sigma})
+8 ln\left(\frac{b (r/r_{\Sigma})-K}{b
-9}\right)\Biggr\}.
\end{equation}

Finally, solving the surface equations for this model, $m$ and $\nu$ are
completely determined and all physical variables can thereby be calculated.
Besides the intrinsic physical interest of the equation of state of this
``seed'' model mentioned before, it is interesting because of the fact that
the static limit of the model
(unlike the previous ones) is ``unstable'', in the sense that it requires
a specific value of the gravitational potential at the boundary, namely
$m_{\Sigma}(0)/r_{\Sigma} =3/14$. For
values above (below) this, the sphere starts to collapse (expand).

Figures (7) and (8) display the evolution of velocity ($\omega$) for
different regions of the sphere, and for initial values of $F$
corresponding to values of
$m_{\Sigma}(0)/r_{\Sigma}$, above and below the equilibrium value,
respectively. Figures (9) and (10), represent the evolution of the Weyl
tensor ($W$),
for some internal region and the boundary surface respectively, and
initials values of $F$ corresponding to values of
$m_{\Sigma}(0)/r_{\Sigma}$, above and below equilibrium. Finally,
figures (11) and (12) exhibit the behaviour of the shear ($\sigma$) for
different regions and initial values of $F$ corresponding to values of
$m_{\Sigma}(0)/r_{\Sigma}$, above and
below equilibrium.

We shall comment on these graphics in the next Section.

\section{Conclusions}

A method has been presented, which allows for the description of radiating
selfgravitating relativistic spheres. In its most general form, the
approach incorporates the two limiting
cases of radiation transport (free streaming and diffusion) as well as the
possibility of dealing with anisotropic fluids.

The cornerstone of the algorithm is  an ansatz, based on an specific
definition of the post--quasistatic approximation, namely: considering
different degrees of departure from
equilibrium, the post--quasistatic regime (i.e. the next step after the
quasistatic situation) is defined as that, characterized by metric
functions whose radial dependence is the same
as that of the quasistatic regime. This in turn implies, that the effective
variables defined above, share the same radial dependence as the
correspondig physical variables of the
quasistatic regime. The rationale behind this definition seems intelligible
when it is remembered that in the latter case (the quasistatic)  the
effective variables  share the same
radial dependence as that of the physical variables in the static regime.
Thus, starting with a static configuration, the first ``level'' off
equilibrium, beyond the quasitatic
situation,  is represented by the post--quasistatic regime.

Once the static (``seed'') solution has been selected, then the definition
of the effective variables together with surface equations, allows for
determination of metric functions,
which in turn lead to the full description of physical variables as
functions of the timelike coordinate, for any region of the sphere. In this
process, depending on the kind of matter
and/or the prevailing transport approximation, additional equations of
state and/or transport equations and/or some of the surface variables (e.g.
the luminosity) have to
be specified.

 All physical variables having been found (particularly the energy density
and the radial pressure) then we may, in principle, go to the next step
assuming that the effective variables
now share the same radial dependence as that of the physical variables just
obtained. In this sense the algorithm may be regarded as an iterative
approach. For obvious reasons we have
restrained ourselves to the first step of the process. It remains to be
seen if available physical evidence justifies to go through the
complexities associated with the
``post--post--quasitatic'' approximation.

In order to illustrate the method, and without the pretension of modeling
specific astrophysical scenarios,  we have presented three examples, in the
simplest (adiabatic) case.

In the first model, the profiles of the shear and the Weyl tensor, clearly
illustrate the ``dynamics'' of the model, tending to zero in the static
limit. The fact that these two
quantities vanish in the quasitatic regime (for this specific model),
brings out further their relevance   in the treatment of situations off
equilibrium. On the other hand however,
the velocity profiles show almost no difference between the two regimes.
Deviations from homology contraction due to  relativistic gravitational
effect are also indicated.

The purpose of the second example was to illustrate the implementation of
the algorithm, for anisotropic fluids. The very particular form of the
``seed'' equation of state of this
model, imposses discontinuity (surface tension) of radial pressure at the
boundary. Of course such discontinuity vanishes at the static (or
quasistatic) regime.

Finally, a model based on the Tolman VI solution was presented. This static
solution, as was already mentioned, requires a specific value of
$m_{\Sigma}(0)/r_{\Sigma}$, accordingly any
deviation from this value leads to deviations from the static regime
(observe that the quasitatic regime is incompatible with this solution).
The velocity profiles indicate that all
regions either expand or contract, and therefore, cracking (different
signs of the velocity for different regions of the sphere) will not occur
\cite{cracking}. This is consistent
with the established fact that cracking  only occurs for anisotropic fluids
or isotropic fluids with outgoing radiation in the free streaming
approximation.

Also, the profiles of the Weyl tensor and the shear, clearly diverging from
the initial values as time proceeds and the evolution becomes more and more
``dynamic'', stress once
again their roles in describing departures off equilibrium.

\acknowledgements
WB was benefited from research support from the Consejo de
Investigaci\'on,
Universidad de Oriente, under Grant CI-5-1002-1054/01 and from FONACIT
under grant S1--98003270 to the Universidad de Oriente. NOS acknowledges
financial support of CNPq. LH acknowledges financial assistance under grant
BFM2000-1322 (M.C.T. Spain) and from C\'atedra-FONACIT, under grant
2001001789.

\begin{figure}
\centerline{\epsfxsize=4in\epsfbox{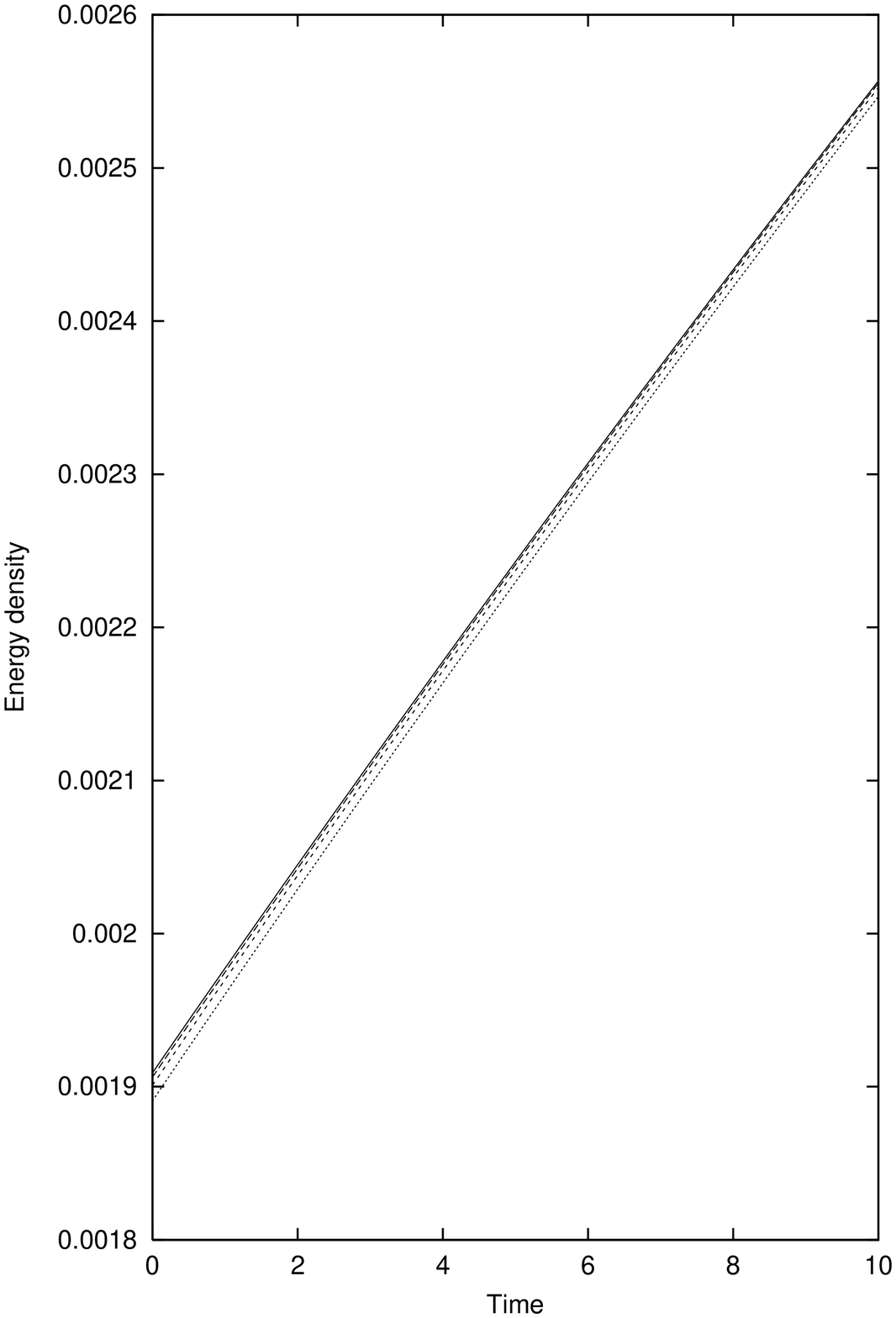}}
\caption{Energy denstity, $\rho\, m(0)^2$, as a function of
(dimenssionless) time ($\alpha$) for the Schwarzschild--type model.
The initial conditions are:   $A(0)=5$, $F(0)=0.6$ and $\Omega(0)=-0.1$.
Curves represent different regions: $r/r_{\Sigma}=0.25$ (continuous line);
$0.50$ (dashed line);
$0.75$ (short--dashed line) and $1.00$ (dotted line).}
\end{figure}
\begin{figure}
\centerline{\epsfxsize=4in\epsfbox{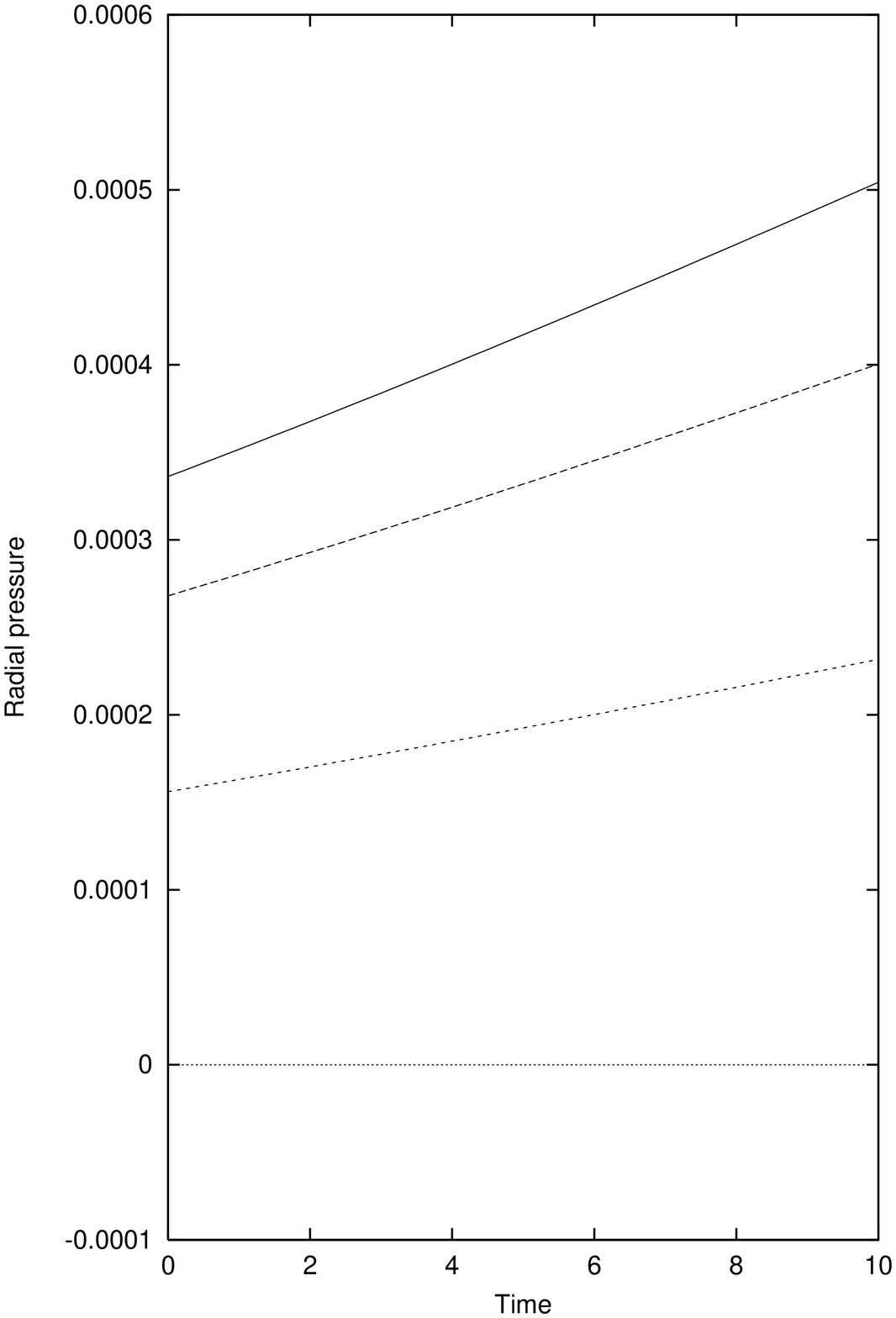}}
\caption{Radial pressure, $P_r \, m(0)^2$, as a function of time for the
Schwarzschild--type model.
The initial conditions are:   $A(0)=5$, $F(0)=0.6$ and $\Omega(0)=-0.1$.
Curves represent different regions: $r/r_{\Sigma}=0.25$ (continuous line);
$0.50$ (dashed line);
$0.75$ (short--dashed line) and $1.00$ (dotted line).}
\end{figure}
\begin{figure}
\centerline{\epsfxsize=4in\epsfbox{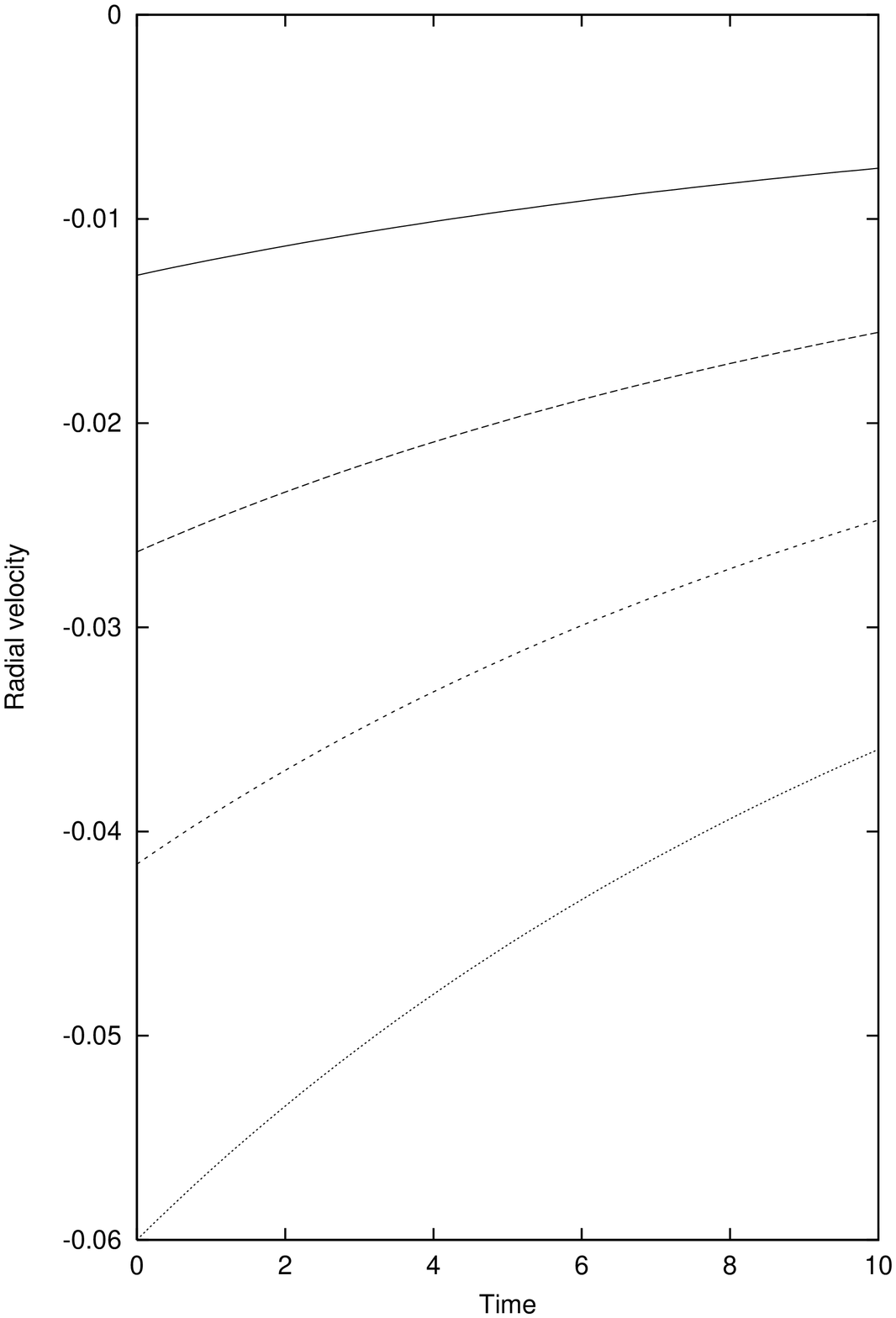}}
\caption{Radial velocity, $\omega$, as a function of time for the
Schwarzschild--type model.
The initial conditions are:   $A(0)=5$, $F(0)=0.6$ and $\Omega(0)=-0.1$.
Curves represent different regions: $r/r_{\Sigma}=0.25$ (continuous line);
$0.50$ (dashed line);
$0.75$ (short--dashed line) and $1.00$ (dotted line).}
\end{figure}
\begin{figure}
\centerline{\epsfxsize=4in\epsfbox{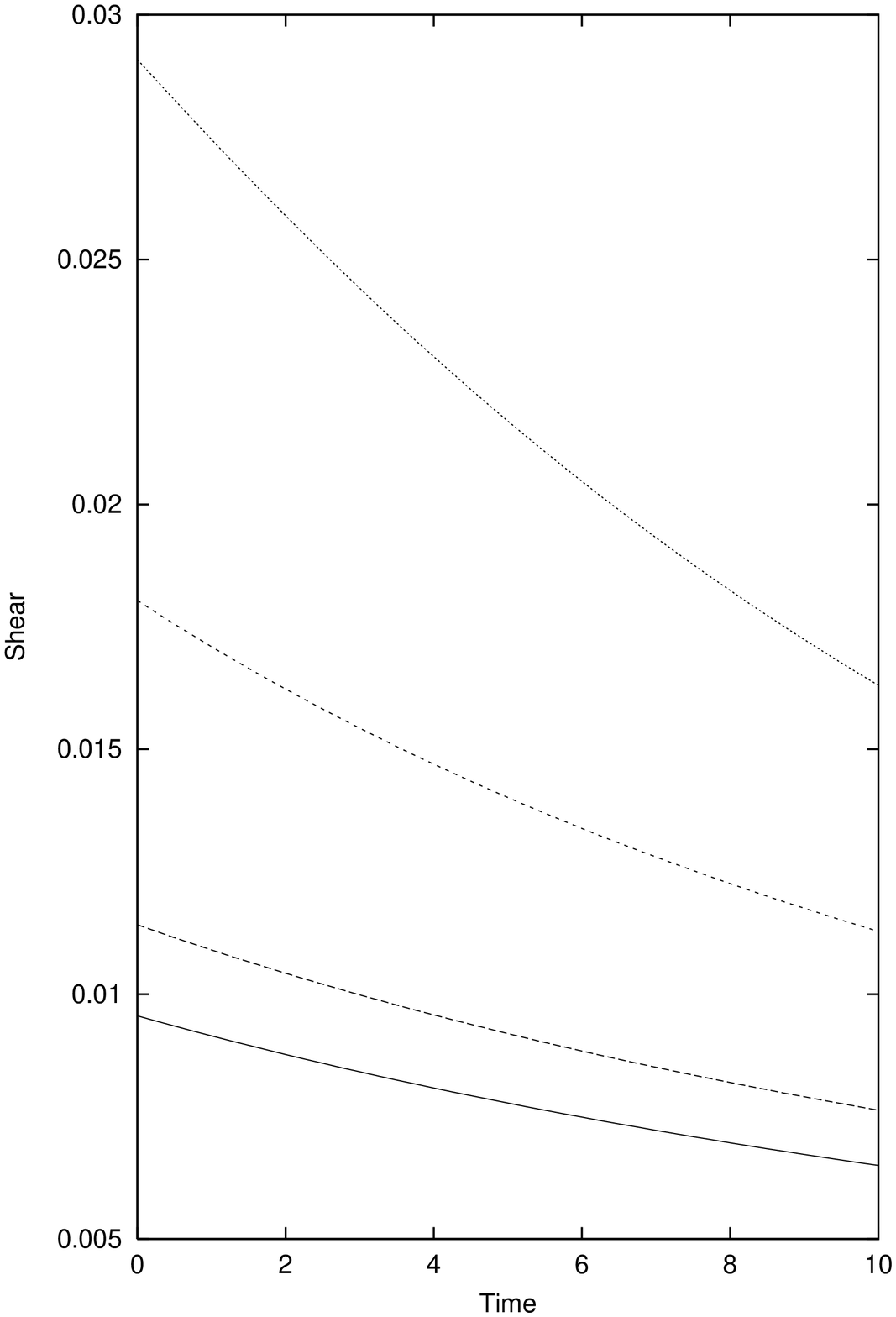}}
\caption{Shear, $\sigma \, m(0)$, as a function of time for the
Schwarzschild--type model.
The initial conditions are:   $A(0)=5$, $F(0)=0.6$ and $\Omega(0)=-0.1$.
Curves represent different regions: $r/r_{\Sigma}=0.25$ (continuous line);
$0.50$ (dashed line);
$0.75$ (short--dashed line) and $1.00$ (dotted line).}
\end{figure}
\begin{figure}
\centerline{\epsfxsize=4in\epsfbox{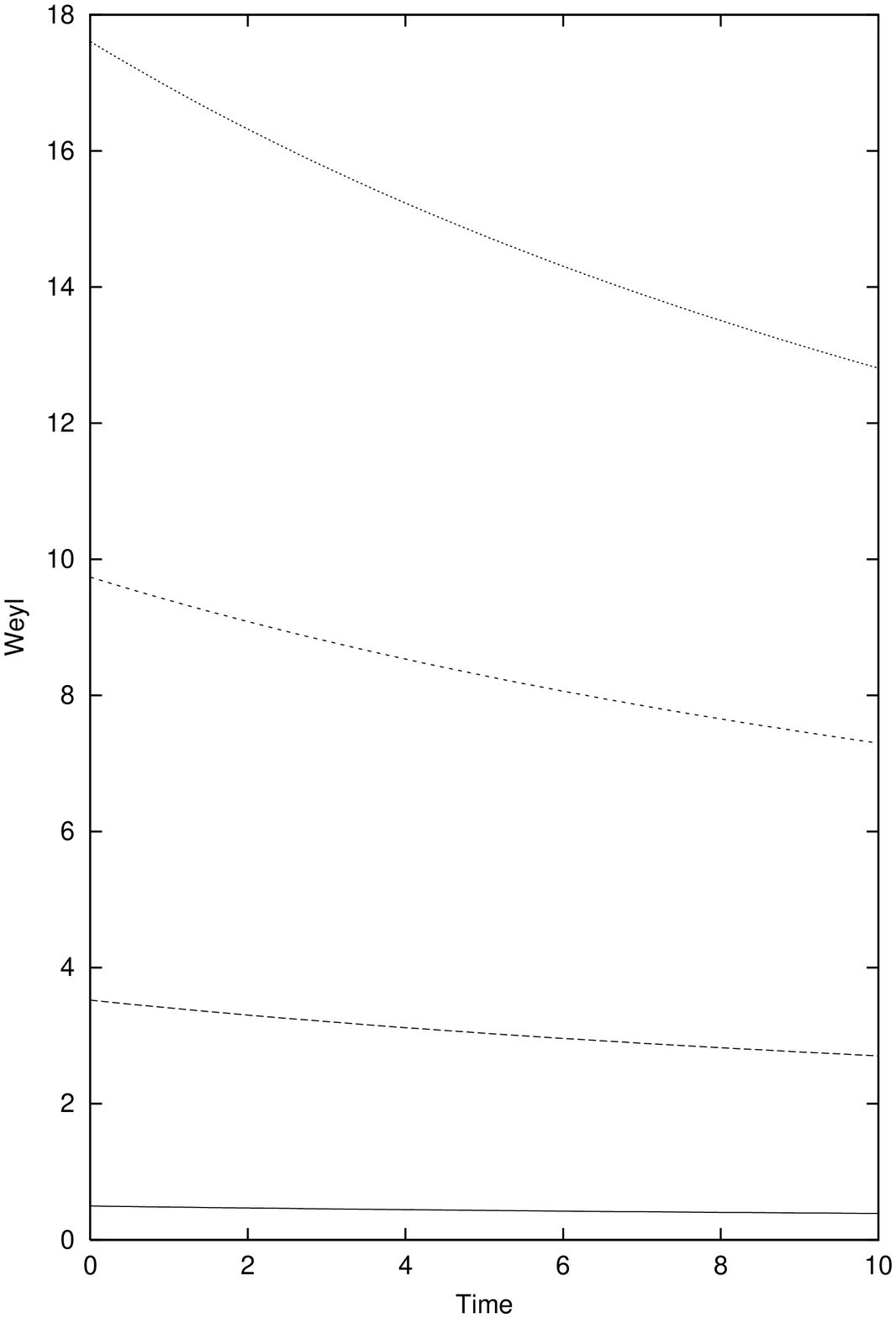}}
\caption{Weyl tensor, $W/m(0)$, as a function of time for the
Schwarzschild--type model.
The initial conditions are:   $A(0)=5$, $F(0)=0.6$ and $\Omega(0)=-0.1$.
Curves represent different regions: $r/r_{\Sigma}=0.25$ (continuous line);
$0.50$ (dashed line);
$0.75$ (short--dashed line) and $1.00$ (dotted line).}
\end{figure}
\begin{figure}
\centerline{\epsfxsize=4in\epsfbox{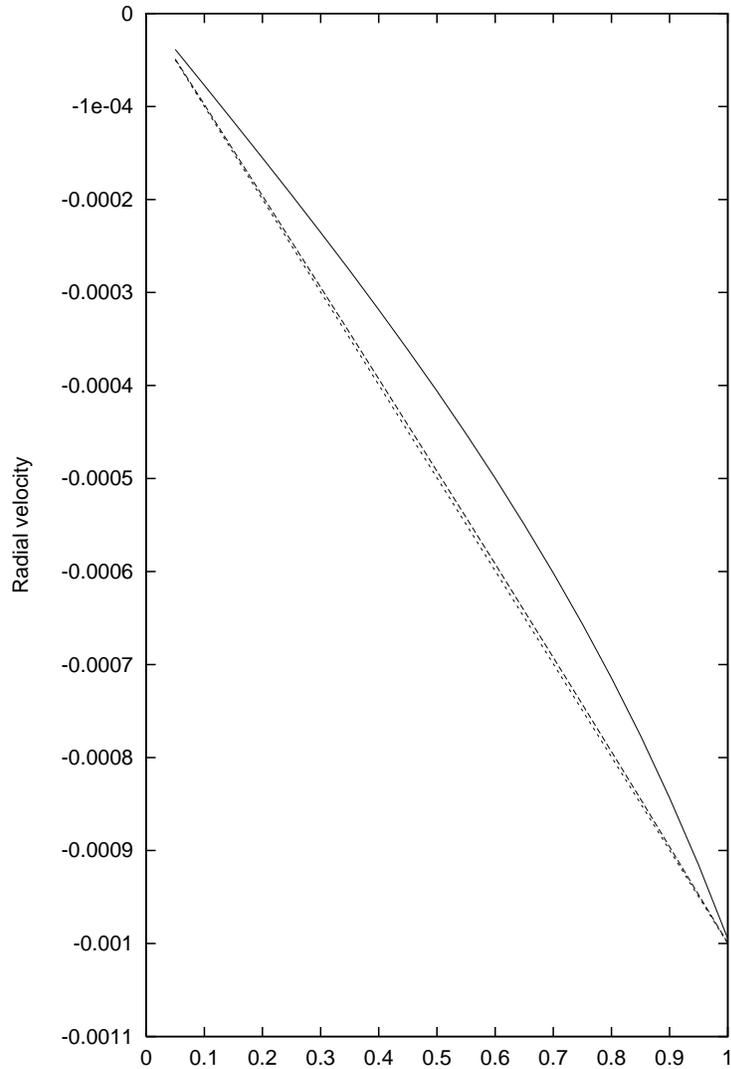}}
\caption{Radial velocity, $\omega$, as a function of $r/r_\Sigma$
for $\alpha=10$. The initial
velocity at the
surface is $-0.001$. Curves represent different values of $F(0)$: $0.6$
(continuous
line); $0.96$ (dashed line) and $0.996$ (short--dashed line).}
\end{figure}
\begin{figure}
\centerline{\epsfxsize=4in\epsfbox{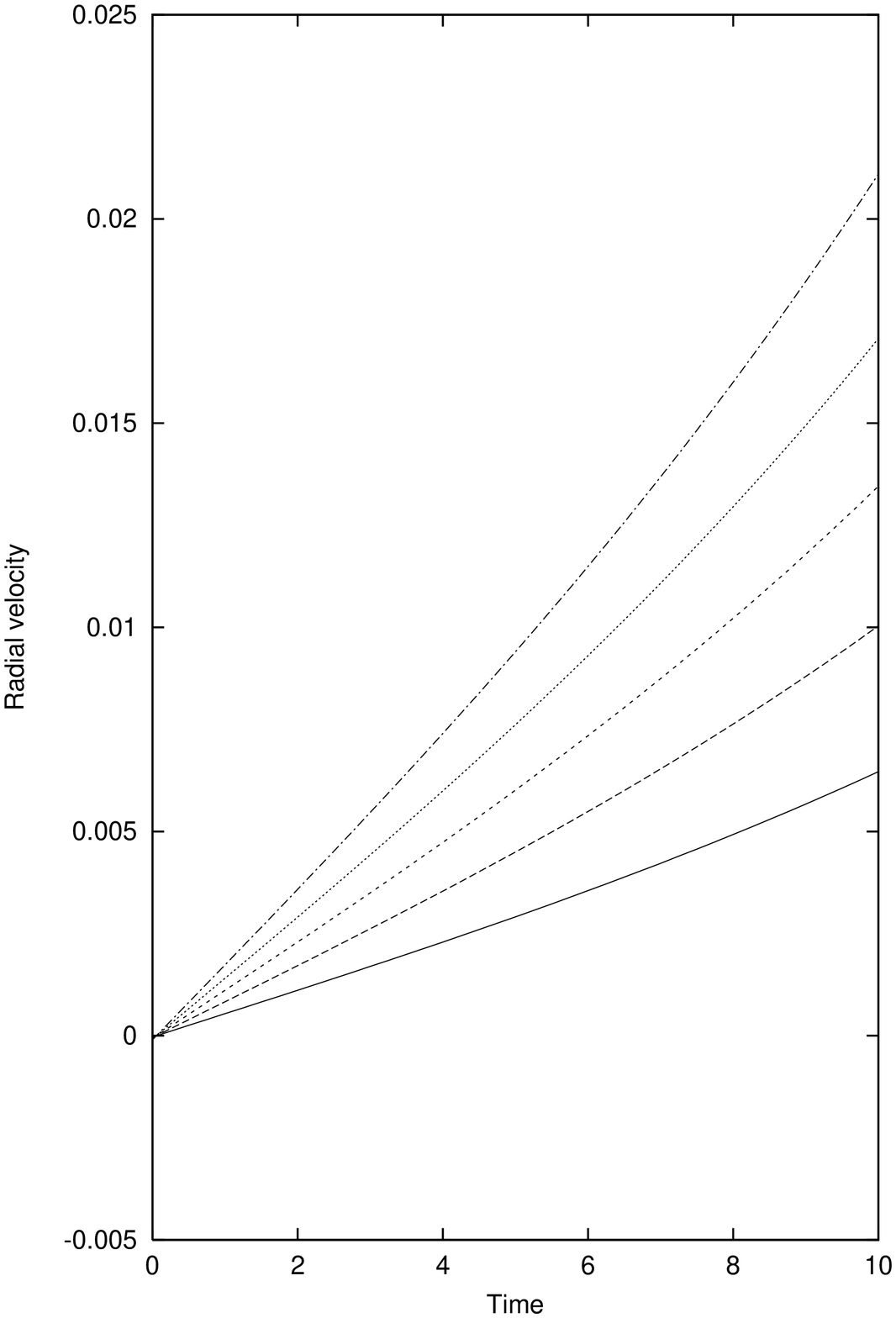}}
\caption{Radial velocity, $\omega$, as a function of time for the Tolman
VI--type model.
The initial conditions are:   $F(0)=0.581428528$ and $\Omega(0)=-0.0001$.
Curves represent different regions: $r/r_{\Sigma}=0.2$ (continuous line); $0.4$
(dashed line);
$0.6$ (short--dashed line); $0.8$ (dotted line)  and $1.0$ (dot--dashed line).}
\end{figure}
\begin{figure}
\centerline{\epsfxsize=4in\epsfbox{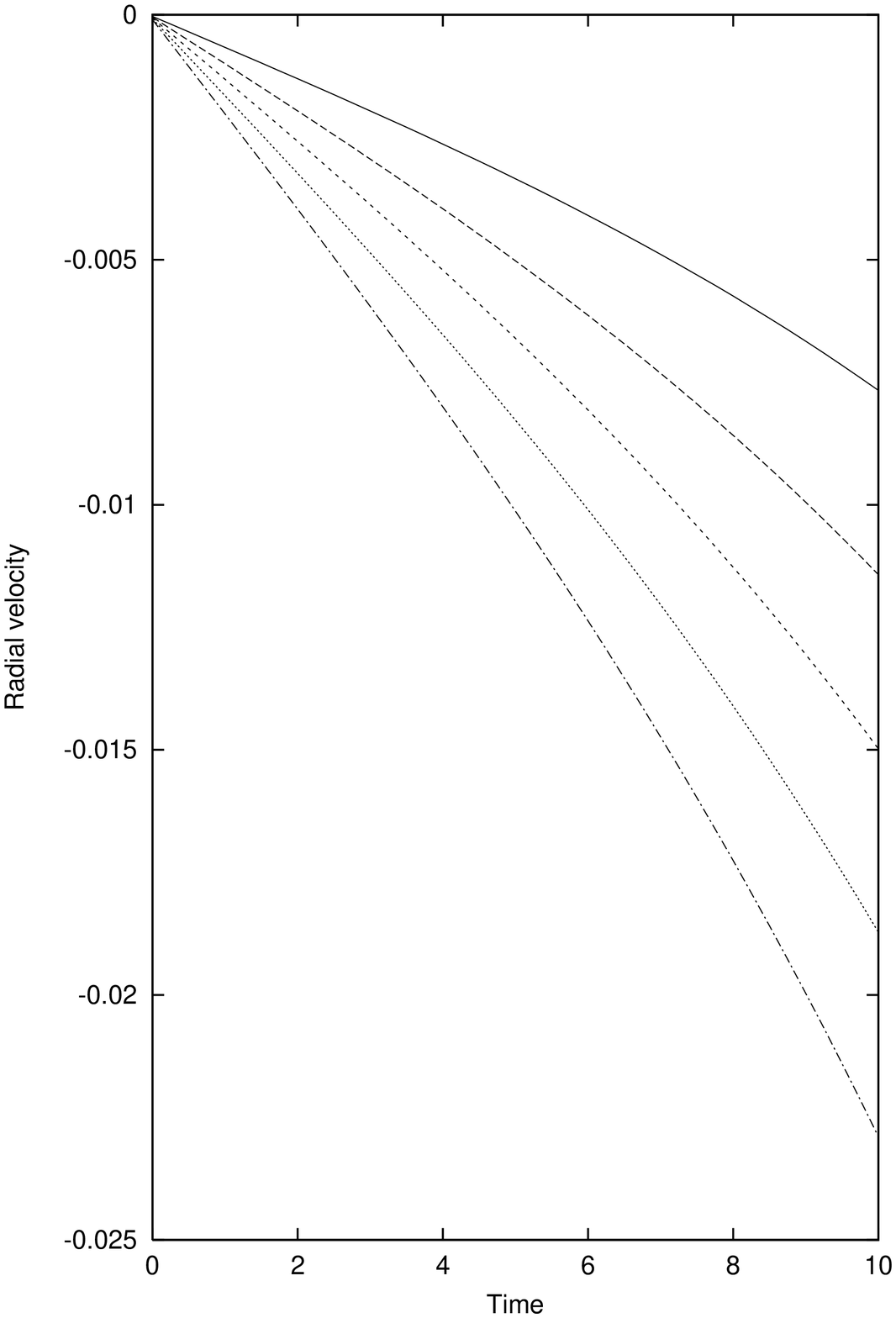}}
\caption{Radial velocity, $\omega$, as a function of time for the
Tolman--type model.
The initial conditions are:   $F(0)=0.561428547$ and $\Omega(0)=-0.0001$.
Curves represent different regions: $r/r_{\Sigma}=0.2$ (continuous line); $0.4$
(dashed line);
$0.6$ (short--dashed line); $0.8$ (dotted line) and $1.0$ (dot--dashed line).}
\end{figure}
\begin{figure}
\centerline{\epsfxsize=4in\epsfbox{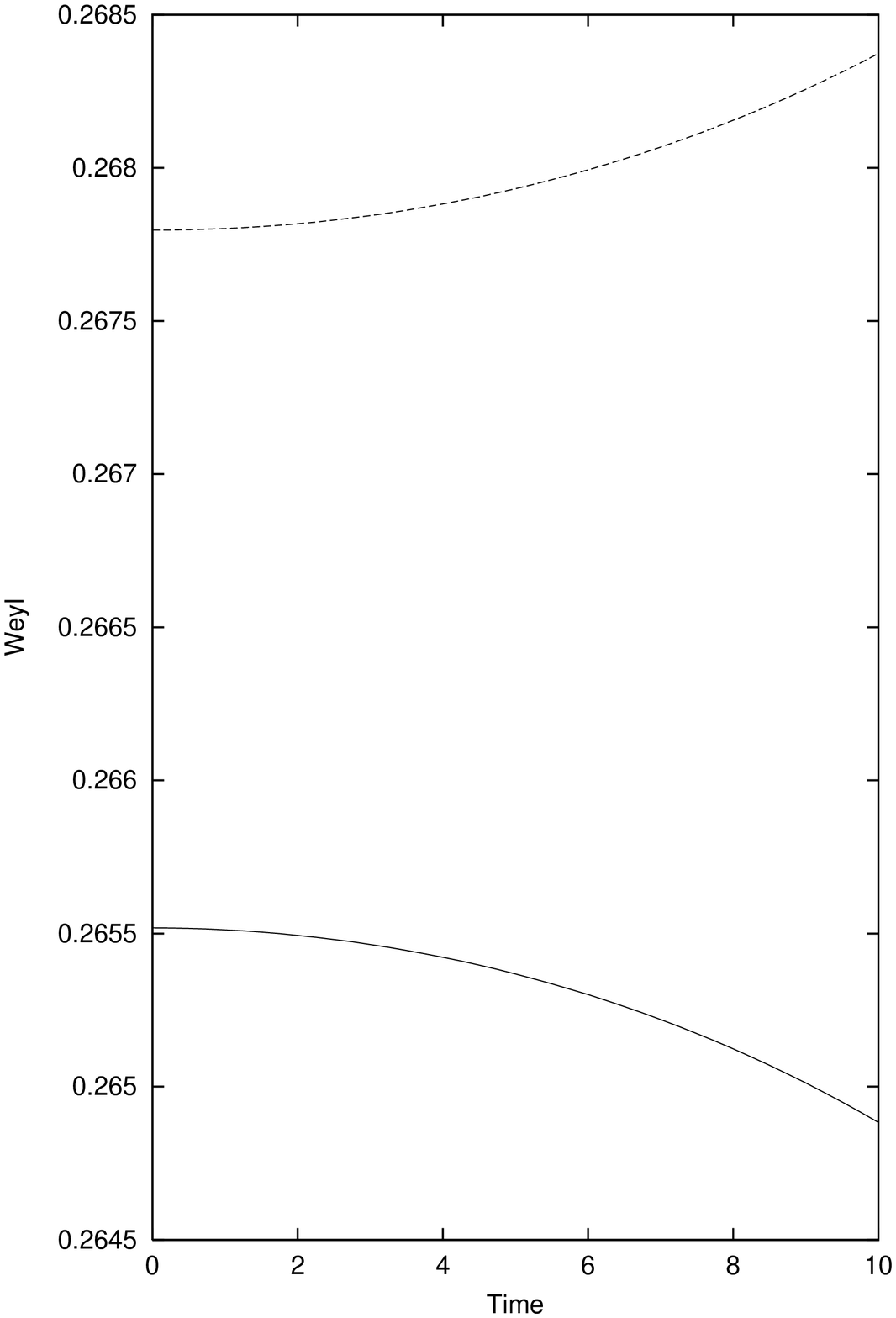}}
\caption{Weyl tensor, $W/m(0)$, at $r/r_{\Sigma}=0.4$ as a function of time
for the Tolman VI--type model.
The initial conditions are:   $F(0)=0.581428528$ (dashed line);
$F(0)=0.561428547$ (continuous line) and $\Omega(0)=-0.0001$.}
\end{figure}
\begin{figure}
\centerline{\epsfxsize=4in\epsfbox{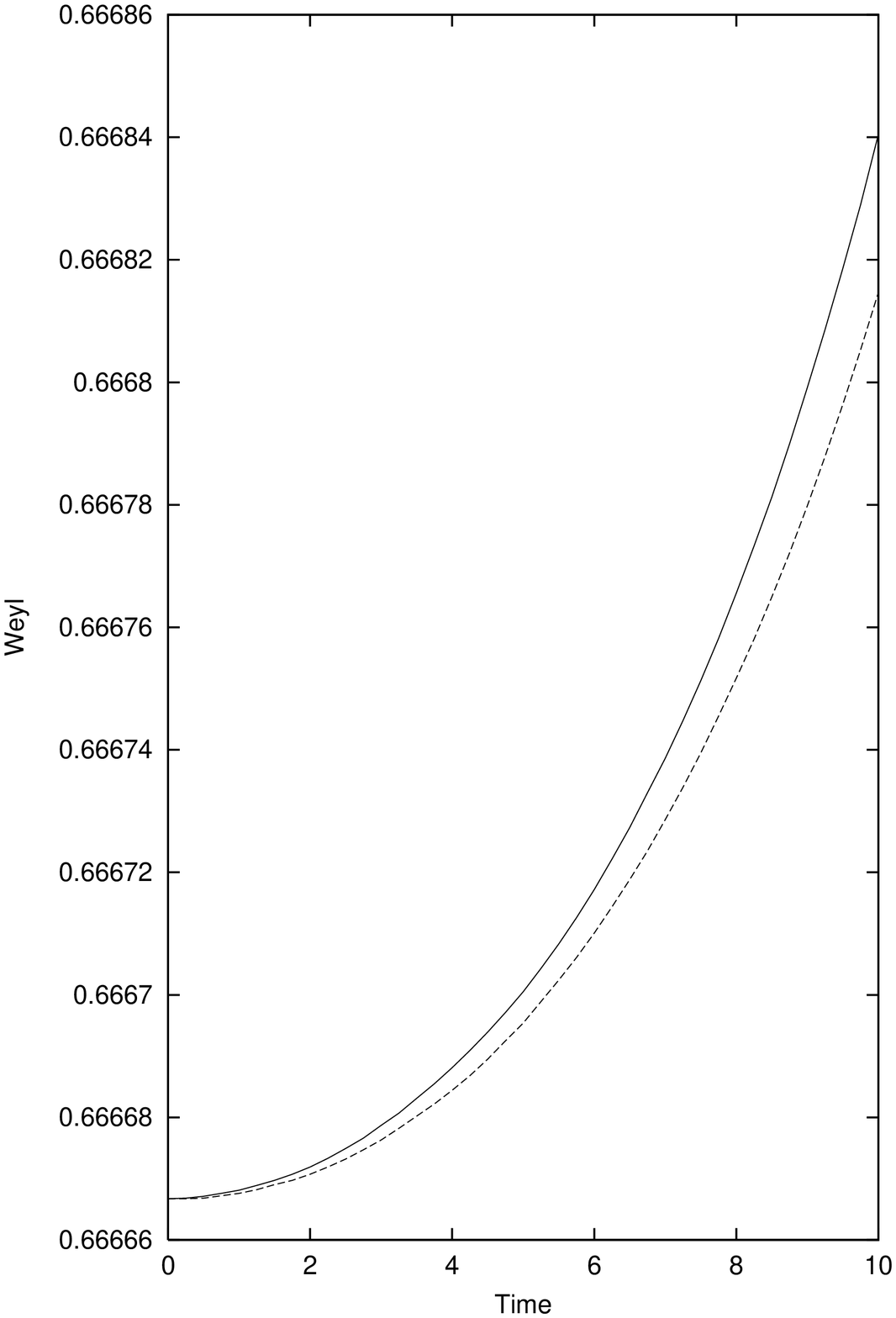}}
\caption{Weyl tensor, $W/m(0)$, at $r/r_{\Sigma}=1.0$ as a function of time
for the Tolman VI--type model.
The initial conditions are:   $F(0)=0.581428528$ (dashed line);
$F(0)=0.561428547$ (continuous line) and $\Omega(0)=-0.0001$.}
\end{figure}
\begin{figure}
\centerline{\epsfxsize=4in\epsfbox{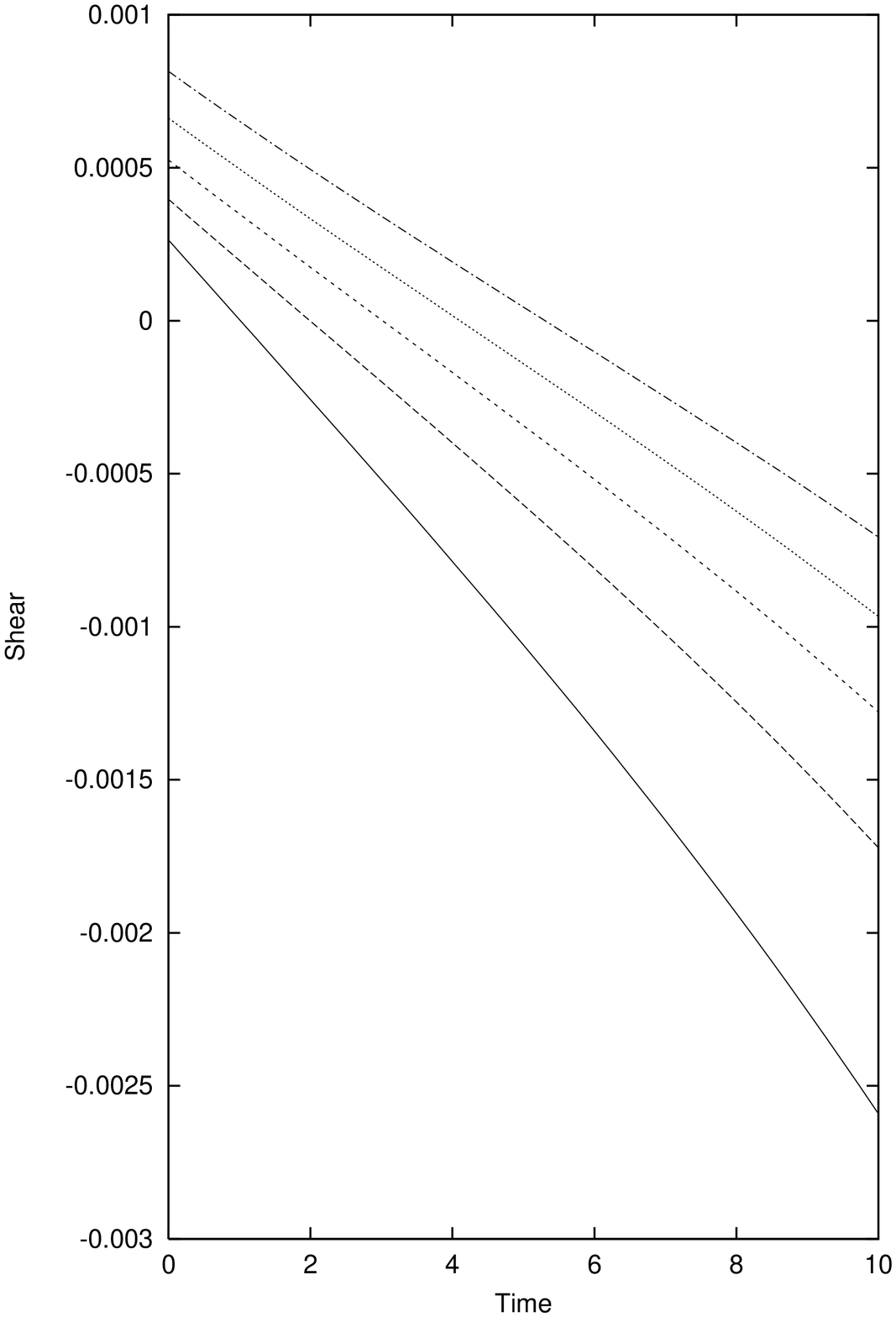}}
\caption{Shear, $\sigma\, m(0)$, as a function of time for the Tolman
VI--type model.
The initial conditions are:   $F(0)=0.581428528$ and $\Omega(0)=-0.0001$.
Curves represent different regions: $r/r_{\Sigma}=0.2$ (continuous line); $0.4$
(dashed line);
$0.6$ (short--dashed line); $0.8$ (dotted line)  and $1.0$ (dot--dashed line).}
\end{figure}
\begin{figure}
\centerline{\epsfxsize=4in\epsfbox{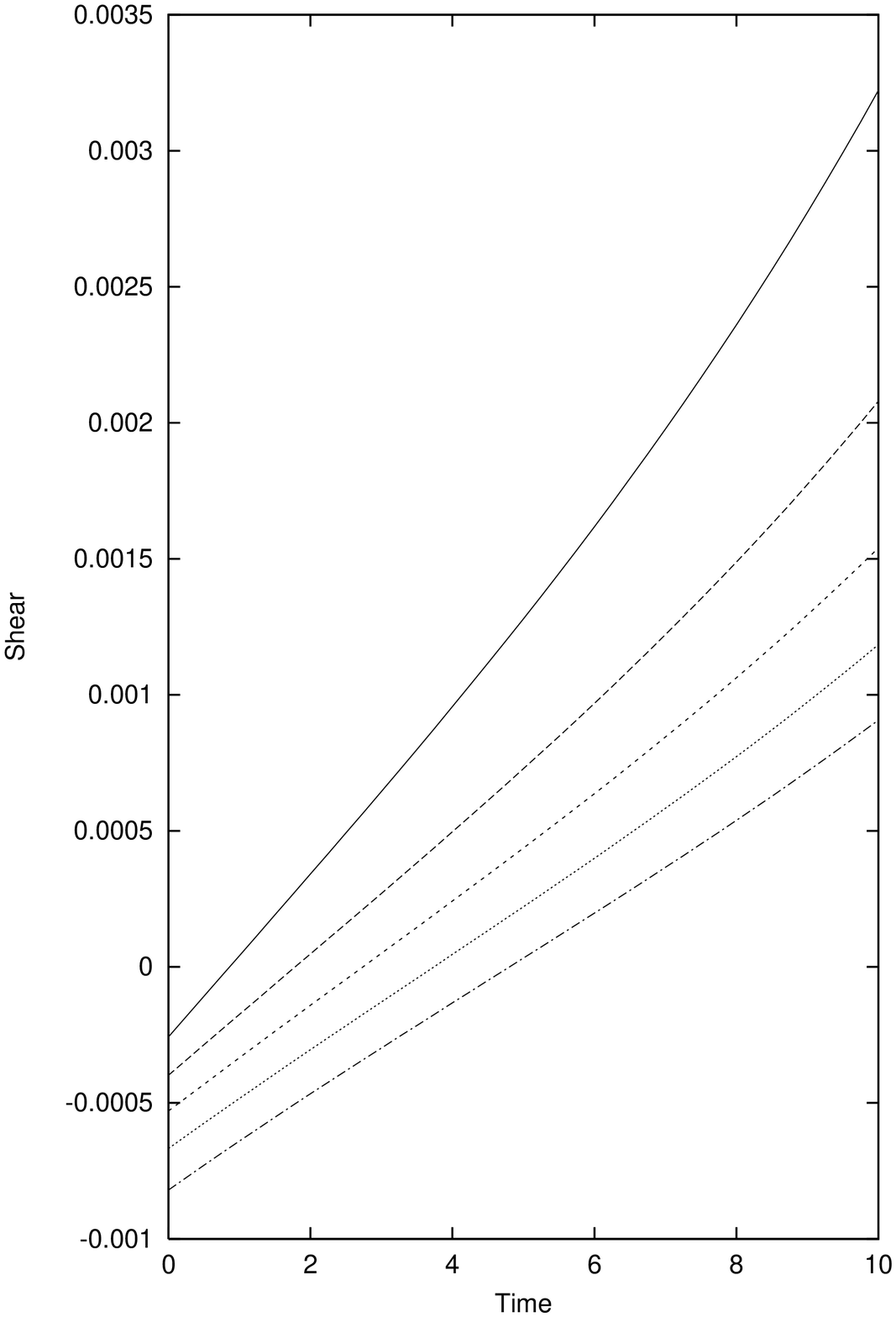}}
\caption{Shear, $\sigma\, m(0)$, as a function of time for the Tolman--type
model.
The initial conditions are:   $F(0)=0.561428547$  and $\Omega(0)=-0.0001$.
Curves represent different regions: $r/r_{\Sigma}=0.2$ (continuous line); $0.4$
(dashed line);
$0.6$ (short--dashed line); $0.8$ (dotted line) and $1.0$ (dot--dashed line).}
\end{figure}
\end{document}